\documentclass[twocolumn]{aastex63}

\usepackage{soul}

\def\heii{\ion{He}{2}~4686~\r{A} }
\def\hheii{\ion{He}{2}~4686~\r{A}}
\newcommand{\sss}{SW~Sex systems }
\newcommand{\ssss}{SW~Sex systems}
\newcommand{\vto}{V380~Oph }
\newcommand{\vvto}{V380~Oph}
\received{}
\revised{}
\accepted{}
\submitjournal{AJ}

\shorttitle{Magnetic accretion in SW Sex Systems}
\shortauthors{Lima et al.}
\graphicspath{{./}{figures/}}

\begin{document}

\title{Search for magnetic accretion in SW~Sextantis systems}

\correspondingauthor{I. J. Lima}
\email{isabel.lima@inpe.br, isabellima01@gmail.com}

\author[0000-0001-6013-1772]{I. J. Lima}
\affiliation{Instituto Nacional de Pesquisas Espaciais (INPE/MCTI), Av. dos Astronautas, 1758, S\~ao Jos\'e dos Campos, SP, Brazil}
\affiliation{Department of Astronomy, University of Washington, Box 351580, Seattle, WA 98195, USA} 

\author[0000-0002-9459-043X]{C. V. Rodrigues}
\affiliation{Instituto Nacional de Pesquisas Espaciais (INPE/MCTI), Av. dos Astronautas, 1758, S\~ao Jos\'e dos Campos, SP, Brazil}

\author[0000-0002-8525-7977]{C. E. Ferreira Lopes}
\affiliation{Instituto Nacional de Pesquisas Espaciais (INPE/MCTI), Av. dos Astronautas, 1758, S\~ao Jos\'e dos Campos, SP, Brazil}

\author[0000-0003-4373-7777]{P. Szkody}
\affiliation{Department of Astronomy, University of Washington, Box 351580, Seattle, WA 98195, USA}

\author[0000-0002-0386-2306]{F. J. Jablonski}
\affiliation{Instituto Nacional de Pesquisas Espaciais (INPE/MCTI), Av. dos Astronautas, 1758, S\~ao Jos\'e dos Campos, SP, Brazil}

\author[0000-0001-6422-9486]{A. S. Oliveira}
\affiliation{IP\&D, Universidade do Vale do Para\'\i ba, 12244-000, S\~ao Jos\'e dos Campos, SP, Brazil}

\author[0000-0003-1949-4621]{K. M. G. Silva}
\affiliation{Gemini Observatory Southern Operations Center - c/o AURA, Casilla 603, La Serena, Chile}

\author[0000-0003-1535-0866]{D. Belloni}
\affiliation{Instituto Nacional de Pesquisas Espaciais (INPE/MCTI), Av. dos Astronautas, 1758, S\~ao Jos\'e dos Campos, SP, Brazil}

\author[0000-0002-0396-8725]{M. S. Palhares}
\affiliation{IP\&D, Universidade do Vale do Para\'\i ba, 12244-000, S\~ao Jos\'e dos Campos, SP, Brazil}

\author{S. Shugarov}
\affiliation{Astronomical Institute of the Slovak Academy of Sciences, Tatransk\'a Lomnica, Slovakia}
\affiliation{Sternberg Astronomical Institute, M.V. Lomonosov Moscow State University, Moscow, Russia}

\author{R. Baptista}
\affiliation{Departamento de F\'\i sica, Universidade Federal de Santa Catarina, Trindade, 88040-900, Florian\'opolis, SC, Brazil}

\author[0000-0002-3817-6402]{L. A. Almeida}
\affiliation{Escola de Ci\^encias e Tecnologia, Universidade Federal do Rio Grande do Norte, Natal, RN 59072-970, Brazil}
\affiliation{Departamento de F\'\i sica, Universidade do Estado do Rio Grande do Norte, Mossoró, RN 59610-210, Brazil}











\begin{abstract}

SW~Sextantis systems are nova-like cataclysmic variables that have unusual spectroscopic properties, which are \replaced{explained}{thought to be caused} by an accretion geometry having part of the mass flux trajectory out of the orbital plane. Accretion onto a magnetic white dwarf is one of the proposed scenarios for these systems. To verify this possibility, we analysed photometric and polarimetric time-series data for a sample of six SW Sex stars. 
We report \replaced{significant}{possible} modulated circular polarization in BO~Cet, SW~Sex, and UU~Aqr with periods of 11.1\deleted{min}, 41.2\deleted{ min,} and 25.7~min, respectively\added{,} and less significant
periodicities for V380~Oph at 22~min and V442~Oph at 19.4~min. We confirm previous results that LS~Peg \replaced{has}{shows} variable circular polarization\replaced{, although}{. However,} we determine a \deleted{different} period of 18.8~min\added{, which is different from the earlier reported value}. We interpret these periods as the spin periods of the white dwarfs. 
Our polarimetric results indicate that 15\% of the SW~Sex systems have direct evidence of magnetic accretion.
We also discuss SW~Sex objects within the perspective of being magnetic systems, considering the latest findings about cataclysmic variables demography, formation and evolution.

\end{abstract}

\keywords{binaries: close — novae, cataclysmic variables — stars: dwarf novae — stars: variables: general — techniques: polarimetric}


\section{Introduction} \label{sec:intro}

SW~Sextantis systems are a class of nova-like cataclysmic variables (CVs). CVs are compact binary systems containing a late-type main-sequence star that is transferring matter onto a white dwarf (WD) via Roche lobe overflow. \replaced{Nova-likes}{Nova-like variables} spend most of the time in a ``high state'' mode, which is \replaced{characterised}{characterized} by high \replaced{accretion}{mass-transfer} rates\added{, $\dot{M}$,} producing steady and bright accretion disks.
\cite{Thorstensen_1991} coined the classification SW Sex for a small group of \replaced{nova-likes}{nova-like variables}\deleted{,} based on their own observations and previous works \citep[e.g.][]{szkody/1990}.
The SW~Sex systems \replaced{have}{are identified by their} spectroscopic characteristics \added{(see below)}\replaced{ that}{, which} challenge explanation within the standard accretion model of CVs.

Initially, SW~Sex objects were associated with eclipsing \replaced{nova-likes}{nova-like variables} despite showing single-peaked emission lines, an unexpected property for high-inclination disks, which normally produce double-peaked lines \added{\citep[e.g.][]{hm86}}. 
Currently, the SW~Sex phenomenon is understood as independent of the system inclination \added{\citep{rodriguez/2007a}}. Additional observational properties of SW~Sex systems are described by \citet{hoard/2003} and summarized below. 
These systems can exhibit absorption in the central part of the Balmer and \replaced{He I}{\ion{He}{1}} emission \added{line} profiles around orbital phase 0.5 (superior conjunction of the secondary star). They also show \replaced{high-ionisation}{high-ionization} emission lines, such as \hheii. The emission line radial velocities are shifted in phase compared to expectations from a simple model of \deleted{the constituents of} the binary system.
The emission lines of SW~Sex systems also show high-velocity S-waves extending up to 4000~km/s,  with maximum blueshift near phase \deleted{$\sim$}0.5.
Interestingly, their orbital periods are typically around 3 to 4~h, just above the period gap. More importantly, \replaced{most of the}{a large number of} CVs in this orbital period range are SW Sex objects \citep{rodriguez/2007b}. This grants SW Sex objects an important \replaced{position for}{role in} the comprehension of CV evolution. An effort to concatenate information about SW Sex systems is the ``The Big List of SW~Sextantis Stars"\footnote{D.~W.\ Hoard's Big List of SW~Sextantis Stars at \url{https://www.dwhoard.com/biglist}. See also \cite{hoard/2003}.} (``The Big List" \replaced{from hereafter}{hereinafter}), which is updated until 2016. 

\added{Some SW~Sex systems show low states, when their brightness drops by \added{a} few magnitudes. This behavior is the defining characteristic of the VY~Scl nova-like systems \citep{Warner_book}. 
Due to those episodic reductions in brightness, they were originally termed \deleted{as} anti-dwarf nova systems.
The recurrence of those low states is irregular and probably associated \replaced{to}{with} a variation of $\dot{M}$. 
In \added{a} high state, the average $\dot{M}$ in SW Sex systems is $\approx 5 \times 10^{-9}\,\rm{M_\odot\,yr^{-1}}$ and not statistically different from the values found for other nova-like systems \citep{puebla2007,Ballouz_2009}.}

Some of the spectroscopic features of SW Sex systems can be explained by the presence of material out of the orbital plane. Several scenarios with this property have been proposed to explain the SW~Sex phenomena  \citep[e.g.][]{hellier/1996,hoard/2003}: an accretion disk wind \citep[e.g.][]{Honeycutt_1986}; a gas stream produced by disk overflow;
a bright/extended hot spot and a flared accretion disk \citep[e.g.][]{dhillon/2013,Tovmassian_2014}; and magnetic accretion \citep[e.g.][]{Williams_1989,casares/1996,rodriguez/2001,hoard/2003}\explain{(One reference was added.)}. Those ideas are not mutually exclusive.

Nova-like systems are commonly classified as non-magnetic CVs \citep[e.g.][]{Dhillon_1996}. 
However, there are claims that the magnetic scenario is more appropriate\deleted{d} to explain the SW~Sex systems \citep[see, for instance,][]{hoard/2003}. The detection of variable circular polarization in some SW~Sex systems is \deleted{a} very strong evidence that magnetic accretion occurs in a fraction of these systems. \replaced{In section \ref{disc:sw_pol} and appendix \ref{app:sw_pol}, we present a compilation of all polarimetric measurements of SW~Sex \added{systems} to date.}{In Appendix~\ref{app:sw_pol}, we present a compilation of all polarimetric measurements of SW~Sex \added{systems} to date, which are discussed in Section~\ref{disc:sw_pol}.}

\added{In magnetic CVs, the accretion flow leaves the binary orbital plane to follow the WD magnetic-field lines creating  one or two accretion magnetic structures that direct matter onto the WD surface. Close to the WD, the matter reaches supersonic velocities due to a quasi-freefall regime and a shock develops. An increase in the density and temperature in the region between the shock front and the WD surface forms the so-called postshock region (PSR).}
\replaced{p}{Optical p}olarized cyclotron emission from \replaced{a}{the} \replaced{post-shock}{PSR} region is a defining characteristic of the CVs termed polars, for which the \replaced{magnetic fields}{magnetic-field intensities} of the \replaced{white dwarfs}{WDs} are \added{in the range} 10 -- 250~MG\replaced{ and}{. In these systems,} the \replaced{circular polarization}{percentage of circularly polarized light in optical wavelengths} is typically  10 -- 30\%. 
Such high values of polarization are reached because the \replaced{post-shock}{PSR} region emission is not diluted by any accretion disk emission, since \added{in polars} the \added{balistic} mass-transfer stream goes directly to the \replaced{shock region near the \replaced{white dwarf}{WD} surface.}{magnetic accretion column and no accretion disk is formed.}
Intermediate polars (IPs), with  magnetic fields \replaced{from}{in the range} 1 -- 10~MG, can also exhibit circular polarization, but \replaced{in}{at} much smaller levels, \replaced{$\sim$1\%~--~2\%}{less than \added{a} few percent\deleted{s}}, since they have truncated accretion disks that \replaced{can also be}{are} an important contribution to the optical emission. 
A compendium of the search for circular polarization in IPs is presented by \citet{Butters_2009}. The presence of a very bright accretion disk in SW~Sex systems may result in even smaller values of optical polarization compared to IPs. 

Other observational properties \added{of the SW Sex systems} are also related to magnetic accretion in CVs\deleted{ and are observed in some SW~Sex systems}: a flux modulation associated with the WD rotation \explain{--- The second reference is new. ---} \citep[e.g.][]{Aviles_2020,KRAur}\replaced{, a periodic variability}{; flux oscillations} in the optical emission-line  wings --- usually \replaced{demonstrated}{referred} as emission-line flaring \citep[e.g.][]{rodriguez/2002}\added{;} and kilosecond quasi-periodic oscillation\added{s} (QPOs) \citep{Patterson_2002}. A review of these properties is given by \citet{hoard/2003}.

\cite{Patterson_2002} suggested that SW Sex systems could also be \replaced{included in the}{classified as} magnetic CVs\deleted{ class}. In a diagram of magnetic momentum, $\mu$, and \deleted{mass-accretion rate,} $\dot{M}$, the SW~Sex systems could be located in the upper right portion with high $\mu$ and high $\dot{M}$. The polars would have similar $\mu$, but smaller $\dot{M}$ and IPs would have smaller $\mu$ (see their Fig.~16). The large values of  $\dot{M}$ in SW~Sex systems would result in small magnetospheres \added{even for intense magnetic fields.}
In this interpretation, SW~Sex stars, which are located mainly above the period gap, would evolve towards smaller periods, and hence smaller  $\dot{M}$, becoming polars below the period gap.

In this paper, we present an observational search for evidence of magnetic accretion in six objects of the SW~Sex class: BO~Cet, SW~Sex, V442~Oph, V380~Oph, LS~Peg, and UU~Aqr. Specifically, we look for periodic variability of flux or polarization that can be associated with the rotation of the WD. Our targets are assigned to the ``Definite'' SW~Sex membership status in \deleted{the} ``The Big List". The observations and reduction procedure are described in Section~\ref{sec_observation}. The adopted methodology for searching for periodicities is presented in Section~\ref{sec_analysis}. A brief literature review and the results for each object are shown in Section~\ref{sec:results}. Section~\ref{sec:discussion} discusses our results in the general scenario of SW~Sex objects as CVs. Finally, the conclusions are given in Section~\ref{sec:conclusions}.

\section{Data description} \label{sec_observation}

We obtained simultaneous photometric and polarimetric time series for a sample of 6 SW~Sex systems. Table~\ref{tab_data} presents a summary of these observations. The observations were performed using the 1.6~m Perkin-Elmer telescope of the \replaced{Observatório}{Observat\'orio} do Pico dos Dias (OPD) \deleted{and the} coupled with the IAGPOL polarimeter \citep{Magalhaes_1996}, which \replaced{is}{was} equipped with a quarter-wave retarder plate and a Savart plate \citep{rodrigues1998}. 
This instrument splits the incident beam of each object in the field of view (FoV) into two orthogonally polarized beams, the so-called ordinary and extraordinary beams.
The observer can use the counts from those beams to obtain differential photometry and/or the polarimetry, as described below. Bias frames and dome flat-field images were collected to correct \added{for} the systematic effects from CCD data. The data reduction was carried out using IRAF\footnote{IRAF (Image Reduction and Analysis Facility) is distributed by the National Optical Astronomy which are operated by the Association of Universities for Research in Astronomy, Inc., under cooperative agreement with the National Science Foundation \citep{Tody/1986, Tody/1993}.} and the {\sc PCCDPACK} package \citep{pereyra/2000,2018Pereyra}.

\begin{deluxetable*}{ccccccccc}
\tablecaption{Summary of the observations.\label{tab_data}}
\tablewidth{0pt}
\tablehead{
\colhead{Object} & \colhead{Date Obs.}  & \colhead{Filter} & \colhead{Mean magnitude} & \colhead{Exp. time} & \colhead{\replaced{Duration}{Time span}} & \colhead{Detector} & \colhead{Telescope} & \colhead{Used in this}\\
\colhead{} & \colhead{} & \colhead{} & \colhead{(mag)} & \colhead{(s)} & \colhead{(h)} & \colhead{} & \colhead{} & \colhead{analysis} 
}
\startdata
{       }  &   2010 Oct 05  & R$_{\rm C}$ & 14.29~$\pm$~0.09  &  60  & 1.34  & \href{http://www.lna.br/opd/instrum/ccd/CCDikonl.html}{IkonL} & OPD & N \\
{       }  &   2010 Oct 06  & V  & 14.16~$\pm$~0.08 &  10  & 3.17  & \href{http://www.lna.br/opd/instrum/ccd/CCDikonl.html}{IkonL} & OPD & Y \\
{       }  &   2010 Oct 11   & V  & 15.13~$\pm$~0.07 &  30  & 1.94  & \href{http://www.lna.br/opd/instrum/ccd/CCDikonl.html}{IkonL} & OPD  & Y\\
{BO Cet}   &   2010 Oct 12  & R$_{\rm C}$ &  14.47~$\pm$~0.10 &  30  & 1.94  & \href{http://www.lna.br/opd/instrum/ccd/CCDikonl.html}{IkonL}  & OPD  & Y\\
{       }  &   2016 Oct 19   & V & 14.53~$\pm$~0.15 &  10  & 1.50  & \href{http://www.lna.br/opd/instrum/ccd/CCDixon.html}{IkonL} & OPD & Y\\
{       }  &   2016 Oct 25  & V &  14.39~$\pm$~0.12 &  10  & 3.87  & \href{http://www.lna.br/opd/instrum/ccd/CCDixon.html}{IkonL} & OPD & N\\   
{       }  & 2019 \replaced{Sept}{Sep} 12 & R$_{\rm C}$ & 14.51~$\pm$~0.07 & 30 & 1.19 & \href{http://www.lna.br/opd/instrum/ccd/CCDixon.html}{Ixon} & OPD & N\\
\hline               
SW Sex         &   2014 Mar 29  & R$_{\rm C}$ & 14.58~$\pm$~0.65  &  40  & 3.29  & \href{http://www.lna.br/opd/instrum/ccd/CCDikonl.html}{IkonL} & OPD & Y\\
\hline
{V442 Oph}   &   2014 Mar 29   & R$_{\rm C}$  & 13.37~$\pm$~0.06 &  20  & 1.44  & \href{http://www.lna.br/opd/instrum/ccd/CCDikonl.html}{IkonL} & OPD & Y   \\
{       }    &   2014 Jul 20 & V  & 13.60~$\pm$~0.05 &  30  & 2.62  & \href{http://www.lna.br/opd/instrum/ccd/CCDikonl.html}{IkonL} & OPD & Y\\
\hline               
{         }     &   2014 Jul 19 & V & 14.51~$\pm$~0.01 &  40                & 2.75  & \href{http://www.lna.br/opd/instrum/ccd/CCDikonl.html}{IkonL} & OPD & Y\tablenotemark{a}\\
{         }    & 2002~--~2016 & V &  14.88~$\pm$~0.16 & \nodata
               & \nodata & \nodata &   SOS/SAS  & Y \\
{V380 Oph}     & 2002~--~2016 & B &  14.86~$\pm$~0.14 & \nodata
               &    \nodata   & \nodata   &   SOS/SAS & Y \\
{         }    & 2002~--~2016 & R$_{\rm C}$ (high state) &  15.05~$\pm$~0.23  & \nodata
               &    \nodata  & \nodata    &   SOS/SAS  & Y \\
{         }    & 2002~--~2016 & R$_{\rm C}$ (low state) &  18.78~$\pm$~0.14  & \nodata    &    \nodata   & \nodata    &   SOS/SAS & Y \\
\hline
{         }   &   2010 Oct 06  & V & 11.82~$\pm$~0.06 &  0.9 & 2.26  & \href{http://www.lna.br/opd/instrum/ccd/CCDikonl.html}{IkonL} & OPD & Y\\
{         }   &   2010 Oct 12  & R$_{\rm C}$ & 11.79~$\pm$~0.08 &  5   & 2.35  & \href{http://www.lna.br/opd/instrum/ccd/CCDikonl.html}{IkonL} & OPD & Y \\
{         }   &   2016 Oct 19  & V & 11.95~$\pm$~0.39 &  2 & 3.65  & \href{http://www.lna.br/opd/instrum/ccd/CCDixon.html}{IkonL} & OPD & Y\\
{LS Peg}      &   2016 Oct 25  & V & 11.93~$\pm$~0.04 &  2   & 0.54  & \href{http://www.lna.br/opd/instrum/ccd/CCDixon.html}{IkonL} & OPD & N \\
{         }   & 2019 \replaced{Sept}{Sep} 09 & V & 11.87~$\pm$~0.05 & 5 & 1.17 & \href{http://www.lna.br/opd/instrum/ccd/CCDixon.html}{Ixon} & OPD & N\\
{         }   & 2019 \replaced{Sept}{Sep} 10 & R$_{\rm C}$ & 11.72~$\pm$~0.10 & 5 & 1.42 & \href{http://www.lna.br/opd/instrum/ccd/CCDixon.html}{Ixon} & OPD & Y\\
{         }  & 2019 \replaced{Sept}{Sep} 12 & R$_{\rm C}$ & 11.54~$\pm$~0.44 & 15 & 2.10 & \href{http://www.lna.br/opd/instrum/ccd/CCDixon.html}{Ixon} & OPD & Y \\
\hline
{       }    &   2009 Aug 21  & V & 12.90~$\pm$~0.14 &  4   & 6.44  & S800 & OPD & N\\
{       }    &   2009 Aug 22   & V  & 12.87~$\pm$~0.09 &  15  & 2.05  & S800 & OPD & N \\
{UU Aqr}     &   2009 Oct 23  & V  & 13.09~$\pm$~0.10 &  15  & 3.02  & S800 & OPD & Y\\
{       }    &   2009 Oct 24  & V  & 13.17~$\pm$~0.32 &  15  & 3.18  & S800 & OPD & N\\
{       }    &   2009 Oct 25  & V   & 13.04~$\pm$~0.08 &  14  & 3.83  & S800 & OPD & Y\\
\enddata
\tablenotetext{a}{Except linear \replaced{polarisation}{polarization} data.}
\end{deluxetable*}

To perform photometry using these polarimetric observations, the total counts from each object \replaced{was}{were} obtained by summing the counts of the ordinary and extraordinary beams. Then, differential photometry was performed by the calculation of the flux ratio between the target objects and a reference star in the FoV that was assumed to be non variable\replaced{ (see Table \ref{tab:ref})}{. The adopted reference stars are shown in Table \ref{tab:ref}}. When possible, we \replaced{adopted}{used} reference stars previously proposed in the literature: \cite{Semena_2013} (LS~Peg) and \citet[HH95]{Henden_1995}. 
We calibrated the instrumental magnitudes using the magnitudes in the NOMAD and USNO-A2.0 catalogues\replaced{(see Table \ref{tab:ref})}{, which are also given in Table \ref{tab:ref}}. 
The estimated accuracy of the conversion from instrumental magnitudes to the calibrated values is around 0.3~mag \citep{Monet_2003}. 

\begin{deluxetable*}{ccccccc}
\tablewidth{0pt} 
\tablecaption{Reference stars used in the magnitude calibration of the targets. \explain{We have used the same decimal plates in all coordinates. } \label{tab:ref}}
\tablehead{
\colhead{} & \multicolumn{6}{c}{Reference star}\\
\cline{2-7}
\colhead{Target} & \colhead{Name} & \colhead{HH95} & \colhead{\replaced{AR}{R.A.}} & \colhead{DEC\added{.}} & \colhead{V} & \colhead{R}\\ 
\colhead{} & \colhead{} & \colhead{} & \colhead{(2000.0)}
& \colhead{(2000.0)} & \colhead{(mag)} &
\colhead{(mag)}
} 
\startdata 
BO Cet & USNO-A2.0 0825-00488714 & \nodata & 02:06:32.13 & $-$02:04:00.3 & 15.1 & 14.1\\
SW Sex & USNO-A2.0 0825-07140676 & SW Sex-2 & 10:15:18.00 & $-$03:07:21.0 & \nodata & 12.93 \\ 
V442 Oph & USNO-A2.0 0737-0410665 & V442 Oph-20 & 17:32:18.03 & $-$16:15:41.2 & 13.97 & 14.53\\
V380 Oph & USNO-A2.0 0960-0317152 & V380 Oph-13 & 17:50:07.14 & $+$06:05:13.8 & 13.9 & \nodata\\
LS Peg & TYC 1134-178-1 & \nodata & 21:52:04.92 & $+$14:05:02.3 & 10.82 & 9.8\\
UU Aqr & USNO-A2.0 0825-19566061 & UU Aqr-5  & 22:09:04.93 & $-$03:46:42.2& 13.80 & \nodata\\
\enddata
\end{deluxetable*}

The polarization was calculated using a differential technique in which the Stokes parameters are estimated from the \replaced{variation}{modulation} of the ratio of the ordinary beam and extraordinary beam counts \explain{--- We removed the "e.g." before the reference, because that reference is the only one that describes the procedure to measure the circular polarization using the IAGPOL instrument. ---}
\citep{rodrigues1998}. 
\added{Specifically, the modulation curve is a function of the Stokes parameters Q, U, and V normalized by the total flux.  The images were acquired in 16 positions of the quarter-wave retarder plate, which are separated by 22.5\degr. To obtain our time series, the waveplate position was continuously increased by this amount.}
One polarization measurement \replaced{was obtained for}{corresponds to} a set of 8 images. We opted for increasing the time resolution at the expense of non-redundant measurements, i.e. to obtain the polarimetric time series we performed a redundant combination of images. Specifically, the first polarimetric point of a series corresponds to the images 1~--~8, the second point corresponds to the images 2~--~9, and so on.
\added{The polarization error was calculated based on the data dispersion around the modulation curve obtained using the estimated Stokes parameters. To evaluate the quality of our polarimetry, we compared this error with the expected value considering the Poisson noise of the source and of the sky, and the detector's gain and readout noise.
The ratio of these two error values varies from 0.5 to 2.0. Therefore, our observational setup provides high-quality data, in the limit of what can be obtained considering the photon and detector noises.} 

\replaced{In SW Sex systems, we expect very low circular polarization values, around tenths of a percent. }
{In SW Sex systems, we expect very low circular polarization values due to the dilution of the PSR region flux by the disk emission. Indeed, the observed circular polarization in SW Sex systems is, with few exceptions, around tenths of a percent (see 
table in Appendix~\ref{app:sw_pol}). An example is V1084~Her, whose circular polarization varies between $-0.5$ and $+0.5$\% \citep{rodriguez/2009}.}
Hence, any improvement in the polarization estimate is worthwhile. With this aim, we implemented an iterative procedure that takes into account the expected value of the sum of the ordinary (and extraordinary) counts in all images used to calculate one polarization point. The method is described in Appendix~\ref{app:k}.

We obtained negligible values for the instrumental polarization, which was estimated using measurements of unpolarized standard stars. \added{Table~\ref{table:standard_stars} shows the average values of the percentage of circular polarization per object and per observational run.} 
\replaced{Hence}{Given the small values and no clear trend in the data}, no correction was applied. 
To convert the position angle of the linear polarization from the instrumental to the equatorial reference frame, we used observations of polarized standard stars\replaced{ and}{. We also} corrected \replaced{the}{all} linear polarization measurements for their positive bias as described by \cite{Vaillancourt_2006}.

\begin{deluxetable}{cccc}
\tablecaption{\explain{This is a new table.} Circular polarization measurements of unpolarized standard stars.
\label{table:standard_stars}}
\tablewidth{0pt}
\tablehead{
\colhead{Run} & \colhead{Object} &  \colhead{$P_c$} & \colhead{Number}
\\
\colhead{} & \colhead{} & \colhead{(\%)} & \colhead{of observations}
}
\startdata
2009 Aug  & HD 154892 & -0.026~$\pm$~0.127 & 1\\ 
2009 Oct  & HD 64299 & 0.007~$\pm$~0.037 & 1\\
2010 Oct  & HD 12021 & 0.01~$\pm$~0.02 & 6\\
          & HD 14069 & 0.02~$\pm$~0.01 & 3\\
2014 Mar  & HD 94851 & -0.01~$\pm$~0.04 & 4\\
          & HD 98161 & 0.12~$\pm$~0.08 & 4\\
2014 Jul  & WD 1620-391 & -0.04~$\pm$~0.01 & 5\\
          & WD 2007-303 & -0.02~$\pm$~0.03 & 5\\
2016 Oct  & HD 14069 & -0.06~$\pm$~0.023 & 1\\
2019 Sep  & HD 154892 & -0.014~$\pm$~0.022 & 1\\
\enddata
\end{deluxetable}

In addition to the OPD observations described above, we also reanalysed the V380~Oph photometry presented in \citet{Shugarov2016} that was taken in BVR$_{\rm C}$ filters and obtained with different telescopes of the Southern Observatory Station (SOS) of Moscow State University and Slovak Academy of Sciences (SAS). The mean magnitude of these data in each filter is presented in Table~\ref{tab_data}.

\section{Data analysis strategy} \label{sec_analysis}

The most characteristic signatures of magnetic accretion in CVs are circularly polarized emission and coherent photometric and/or polarimetric variability associated with the rotation of the WD. Therefore, we performed a search for periodicities in the magnitude, circular polarization, and linear polarization time series of all objects \replaced{of}{in} our sample. 
This search was performed separately for the polarimetric and photometric data because the \replaced{latter}{flux} may show other periodicities that are not related to the WD rotation.
In this section, we describe the common procedures in the data analysis of all objects. 

The Lomb-Scargle \added{(LS)} method \citep{Lomb_1976, Scargle_1982} was used to search for periodicities. We adopted a maximum frequency of ~240 ~d$^{-1}$ ($6$ min), while 
the minimum frequency was adjusted for each star, having typical values around 14~d$^{-1}$ (100~min).
The adopted frequency grid is composed of at least $10^5$ elements, which is adequate to recover the correct period as discussed by, e.g. \citet{FerreiraLopes_2018}.
\deleted{We estimated the period error by the full width at half maximum (FWHM) of a Gaussian adjusted to the region of the frequency maximum. If the region has many peaks caused, for example, by aliasing, we fit a Gaussian to the peaks envelope.}

To minimize spurious peaks in the power spectra of the photometric data, the \replaced{Lomb-Scargle}{LS} method was applied after the subtraction of the mean magnitude of each night. This procedure also removed possible long-term modulations present in the time series. Since the mean polarization is very small, no mean level was removed from the linear and circular polarization data.
For some objects, we filtered low-frequency signals, such as the orbital period or any other known/detected low-frequency period of the system.

\added{To have a realistic estimate of the uncertainty of a signal detected with the LS periodogram,
we scrambled the y-coordinates of the correspondent time series, without repetition.
Then we injected a signal
centered at the detected frequency, but randomly distributed around a few frequency resolution
elements and with a random reference epoch between $0-2\pi$. The amplitude of the injected signal
was the amplitude derived from the LS periodogram. 
We ran this simulation 1,000 times. For each realization, it was calculated the difference between injected and recovered periods.
The standard deviation of the difference, obtained from the median absolute deviation (MAD), is
quoted as the uncertainty of the periods found in our analysis.} 

\added{We compared the errors provided by the above simulations with two analytical estimates. From eq.~2 of \citet{2020ApJ...902...24D} \citep[see also][]{1971AJ.....76..544L}, the period error, $\sigma_P$, is:}

\added{\begin{equation}\label{eq_1}
  \sigma_P = \frac{\sigma P^2}{\pi T A} \sqrt{\frac{6}{N}},  
\end{equation}}

\added{\noindent where $P$ is the period, $\sigma$ is the standard deviation of the data, $T$ is the temporal baseline of the time series, $A$ is the amplitude of signal at $P$, and $N$ is the number of points in the time series. This error estimate is a good approximation for time series in which the counting of the cycles is not lost, as in a continuous time series. However, when the data collected are on different nights, which can be years apart, there is an additional error source and it can be shown that a more adequate estimate is:}

\added{\begin{equation}\label{eq_sigma_P}
  \sigma_P = \frac{\sigma P}{2\pi A} \sqrt{\frac{2}{N}}.  
\end{equation}}

\added{\noindent In this expression, the cycle uncertainty was expressed by considering $T = P$. The uncertainty of the phase, corresponding to a factor equals to 1/12 in a uniform distribution, was not considered in Equation~\ref{eq_sigma_P} in spite of being considered in Equation~\ref{eq_1}. The period errors from the simulation are of the same order of magnitude as the values obtained using Equation~\ref{eq_sigma_P} for data spanning one or more nights. Therefore, this equation can be used as a quick estimate of period errors. }

The emission from a likely magnetic accretion structure would be strongly diluted by the very bright accretion disk of \deleted{ok} SW~Sex systems. Therefore, if present, the circular polarization and the amplitude of the spin-modulated emission must be very small, as is indeed observed in the \added{known} polarized SW~Sex systems (see \replaced{Introduction}{Appendix~\ref{app:sw_pol}}). 
In order to avoid the detection of spurious signals but not lose real features, we adopted a set of criteria to check the reliability of the data sets, as described below.

\begin{figure*}[htb]
\plotone{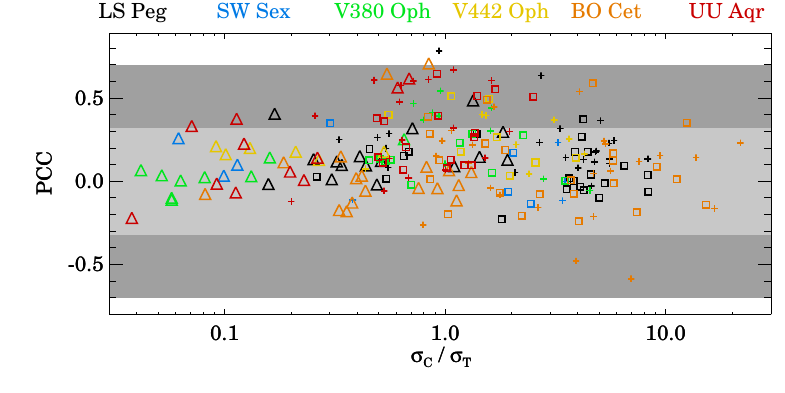}
\caption{The Pearson correlation coefficient (PCC) as a function of the ratio of the standard deviation of the measurements of a given comparison star ($\sigma_{\mathrm C}$) relative to the value of the target ($\sigma_{\mathrm T}$). The triangles represent the photometric data, the plus signs refer to the circular polarimetry, and the squares denote linear polarimetry data. Different colours are used to distinguish each SW Sex star of our sample. We compare each target with more than one field star. This procedure results in a number of points having the same color and symbol. The points inside the light grey region do not present correlation, as explained in \added{the} text. The dark grey region contains data sets that should be inspected  individually to verify the presence of correlation. \label{fig_coeffPerson}}
\end{figure*}

We verified if the magnitude and polarization time series of our targets are correlated with those of the field stars. Correlations can be caused by instrumental effects, such as the usage of a rotating optical element in the observation procedure, that could cause some \replaced{additional}{artificial} modulation in the observed counts. To evaluate the presence of such a correlation, we initially calculated the Pearson correlation coefficient (PCC) between target and field-stars time series. Fig.~\ref{fig_coeffPerson} shows the PCC as a function of the ratio of the standard deviations of the field-star time series to those of the target stars. For most targets, we tested the data using more than one field star. Hence, Fig.~\ref{fig_coeffPerson} can have more than one point with a given symbol and colour. The points inside the light grey region have $|PCC|< 0.32$ and are considered as non-correlated data sets, while the dark grey bands indicate possible correlations ($0.32~<~|PCC|~<~0.70$). The limit of $|PCC|~=~0.32$ was not an {\it a priori} assumed value. It results instead from the complete analysis of the correlation between the flux and polarization of field stars and target systems, which were inspected individually. Only a fraction of the data sets in the dark grey region \replaced{was}{were} removed from the analysis.  

The significance of a period was also assessed through the comparison of the science object power spectrum with those of field stars on a nightly basis. We removed from the analysis the data sets that showed similar peaks in science and field-stars periodograms. The last column of Table \ref{tab_data} states if a given data set was \replaced{removed from}{included in} the analysis as a result of the procedure just described.

The small number of targets allow\deleted{s} us to perform an individual and careful analysis of each source to take into account its peculiarities. Therefore, some adjustments to the \added{above} procedure \deleted{above} were performed when required. The detailed analysis and results for each source \replaced{is}{are} described in Section~\ref{sec:results}.

\section{Results and analysis} \label{sec:results}

In the following sections, we present an overview of previous findings, our results on a period search, and our analysis of the presence of magnetic accretion in each object of our sample. Table~\ref{table:periods} presents the main periods found in the literature as well as those found in this paper.

\begin{deluxetable*}{cccccc}
\tabletypesize{\scriptsize}
\tablewidth{0pt} 
\tablecaption{Claimed periods from \added{the} literature and from our analysis\deleted{ for our SW~Sex sample}. We indicate our periods having small confidence levels with a trailing question mark. \explain{We have made some small changes in this table. No changes have been made in column 2.} \label{table:periods}}
\tablehead{
\colhead{Object} & \colhead{Period} & \colhead{Data \added{type}} & \colhead{Interpretation} & \colhead{Reference}\\
} 
\startdata 
{      }& 3.36 h & Photometry & P$_{\rm orb}$ & 1 \\
{      }& 19.9~$\pm$~0.9 min & Spectroscopy (H$\alpha$ RV) & P$_{\rm spin}$ & 2\\
{      }& 19.6~$\pm$~0.9 min & Spectroscopy (H$\alpha$ EW) & P$_{\rm spin}$ & 2\\
\\
{      }& 50.9~$\pm$~\replaced{6.3}{0.5}~min & Photometry & --- & This work\\
{BO Cet}& 19.7~min & Photometry & ---  & This work\\
{      }& 15.3~min & Photometry & --- & This work\\
{      }& 11.1~$\pm$~\replaced{0.3}{0.08}~min & Circular Pol. & P$_{\rm spin}$ & This work\\
{      }& 9~min & Circular Pol. & P$_{\rm beat}$ (50.9 - 11.1) & This work\\
{      }& 14~min & Circular Pol. & P$_{\rm beat}$ (50.9 - 11.1) & This work\\
\hline
{      } & 3.24 h & Photometry & P$_{\rm orb}$ & 3 \\
\\
{SW Sex}& 22.6~$\pm$~\replaced{3.8}{1.4}~min & Photometry & P$_{\rm spin}$/2 & This work\\
{      }& 41.2~$\pm$~\replaced{7.7}{8.5}~min & Circular Pol. & P$_{\rm spin}$ & This work\\
\hline
{      }& 2.98 h & Spectroscopy (RV) & P$_{\rm orb}$ & 4 \\
{      }& 4.37 d & Photometry & P$_{W}$ & 5\\
{      }& 2.90 h & Photometry & P$_{\rm SH-}$ & 5\\
{      }& 16.66 min & Photometry & QPOs & 5\\
{V442 Oph}& 16.0 min & Photometry & QPOs? & 5\\
{      }& 19.5 min & Photometry & QPOs? & 5\\
\\
{      }& 12.4~$\pm$~\replaced{0.5}{0.09}~min & Photometry & --- & This work\\
{      }& 19.4~$\pm$~\replaced{1.7}{0.4}~min (?) & Circular Pol. & P$_{\rm spin}$ & This work\\
{      }& 18.3~$\pm$~1.1~min (?)
{      }& Linear Pol. & P$_{\rm spin}$  & This work\\
\hline
{      } & 3.70 h & Spectroscopy (H$\alpha$ RV) & P$_{\rm orb}$ & 2 \\
{      }& 4.51 d & Photometry & P$_{W}$  & 6 \\
{      }& 3.56 h & Photometry & P$_{\rm SH-}$  & 6 \\
{      }& 46.7~$\pm$~0.1~min & Spectroscopy (H$\alpha$ EW) & P$_{\rm spin}$ & 2\\
{ V380 Oph }\\
{      }& 47.4~$\pm$~\replaced{4.3}{4.9}~min & Photometry (V band) & --- & This work\\
{      }& 12.4~$\pm$\added{0.66}~min & Photometry (B band) & P$_{\rm spin}$?  & This work\\
{      }& 22.0~$\pm$~\replaced{2.2}{1.2} min (?) & Circular Pol. & P$_{\rm spin}$?  & This work\\
{      }& 12.7~$\pm$~\added{0.4}~ min (?) & Circular Pol. & P$_{\rm spin}$?  & This work\\
\hline
{      }& 4.2 h & Spectroscopy (H$\alpha$ RV) & P$_{\rm orb}$ & 7,8 \\
{      }& 20.7~$\pm$~0.3 min  & Photometry & QPOs  & 7 \\
{      }& 19 min & Photometry & P$_{\rm spin}$? & 9\\
{      }& 16.5~$\pm$~2 min & Photometry & --- & 10\\
{      }& 19 min & Photometry & QPOs &  11\\
{      }& $\sim$20 min & Spectroscopy & QPOs &  12\\
{      }& 33.5~$\pm$~2.2 min min & Spectroscopy & P$_{\rm beat}$ & 13 \\
{LS Peg}& 29.6~$\pm$~1.8 min & Circular Pol. & P$_{\rm spin}$ & 13\\
{      }& 30.9~$\pm$~0.3 min\tablenotemark{a} & X-rays & P$_{\rm spin}$  & 14\\
\\
{      }& 21.0~$\pm$~\replaced{1.7}{1.2}~min & Photometry & P$_{\rm beat}$ (P$_{\rm orb}$ - P$_{\rm spin}$) & This work\\
{      }& 16.8~min & Photometry & --- & This work\\
{      }& 19.3~min & Photometry & --- & This work\\
{      }& 24.2~min & Photometry & --- & This work\\
{      }& 18.8~$\pm$~\replaced{1.0}{0.005}~min & Circular Pol. & P$_{\rm spin}$  & This work\\
{      }& 11.5~$\pm$~0.1~min & Linear Pol. & P$_{\rm spin}$/2  & This work\\
\hline
{      } & 3.9 h & Photometry & P$_{\rm orb}$ & 15\\
{      }& 4.2 h & Photometry &  P$_{\rm SH+}$ & 16\\
{UU Aqr}& 4 d & Photometry & --- & 17 \\ 
\\
{      }& 54.4~$\pm$~\replaced{4.6}{0.5}~min & Photometry & 2 P$_{\rm spin}$ & This work\\
{      }& 25.7~$\pm$~\replaced{0.5}{0.23}~min & Circular Pol. & P$_{\rm spin}$ & This work\\
\enddata
\tablenotetext{a}{This X-ray periodicity was not confirmed by \replaced{posterior}{later} XMM observations \citep{Ramsay_2008}.}
\tablecomments{\added{Symbols and acronyms used in this table are as follows.} P$_{\rm orb}$: orbital period; P$_{\rm spin}$: spin period; P$_{\rm beat}$: beat period
\added{(the two periods causing the beat are presented between brackets)};
P$_{\rm SH+}$: positive superhump period; P$_{\rm SH-}$: negative superhump period; P$_{W}$: Disk wobble period\added{; RV: Radial velocity; EW: Equivalent width.}}
\tablerefs{(1)~\cite{Bruch2017},
(2)~\cite{rodriguez/2007b},
(3)~\cite{Groot_2001},
(4)~\cite{Hoard_2000},
(5)~\cite{Patterson_2002},
(6)~\cite{Shugarov_2005},
(7)~\cite{Taylor_1999},
(8)~\cite{Martinez_1999},
(9)~\cite{Garnavich_1992}, 
(10)~\cite{Szkody_1994}, 
(11)~\cite{Szkody_2001}, 
(12)~\cite{Szkody_1997}, 
(13)~\cite{rodriguez/2001}, 
(14)~\cite{baskill/2005},   
(15)~\cite{Baptista_1996}, 
(16)~\cite{Patterson_2005}, 
(17)~\cite{Bruch_2019}}
\end{deluxetable*}

\subsection{BO~Cet}  \label{results_bocet}

BO~Cet was \replaced{initially}{first} classified as a nova-like CV with V~$\approx$~14~--~15~mag by \cite{Downes_2005}.
Its spectrum displays the typical single-peaked H~I emission lines of SW~Sex systems, along with double-peaked \ion{He}{1} \deleted{emission lines} and relatively strong \heii emission \added{lines} \citep{zwitter_1995}.
BO~Cet was classified as a non-eclipsing SW~Sex star by \cite{rodriguez/2007a}, who published a spectroscopic study of this source. 
The H$\alpha$ trailed \replaced{spectrum shows}{spectra show} an S-wave with an amplitude of $\sim$300~km~s$^{-1}$, with the bluest velocity and minimum intensity both occurring at phase 0.5, as well as a high-velocity S-wave component reaching $\sim$2000~km~s$^{-1}$.
This high-velocity component exhibits periodic emission-line flaring at a period of 19.9~$\pm$~0.9~min, which is interpreted as the spin period of a magnetic WD. Its orbital period is estimated as 0.13983~d by \cite{Bruch2017} based on the Center for Backyard Astrophysics (CBA) data. The only hitherto formally published photometric time series of BO~Cet is by \cite{Bruch2017}. His data sets show strong flickering, and the orbital modulation is not always discernible.
\added{AAVSO\footnote{See \url{https://www.aavso.org/LCGv2/}} data taken over 2020 show an interesting variability pattern: BO~Cet magnitudes oscillate between 13.6 and 16.2~mag in an approximate cadence of 2 weeks. This behavior suggests a possible Z~Cam classification as already mentioned by \cite{Taichi_2018}.}

We observed BO~Cet in \added{the} V and R$_{\rm C}$ bands during three runs in 2010, 2016 and 2019 (see Table~\ref{tab_data}) with exposure times between 10 \deleted{s} and 60~s. Our data do not show any significant variation in the average magnitude in the R$_{\rm C}$ band, but the V band average varied by around 1~mag (see Table~\ref{tab_data}).

Our data are not suitable to refine the orbital period, since the individual data sets span time intervals comparable to or smaller than the \added{suggested} orbital period, and the long term data sampling introduces a huge aliasing structure in the power spectrum. Moreover, \citet{Bruch2017} shows that an orbital modulation is not always clearly present in BO~Cet photometry. 

\begin{figure*}
\gridline{\fig{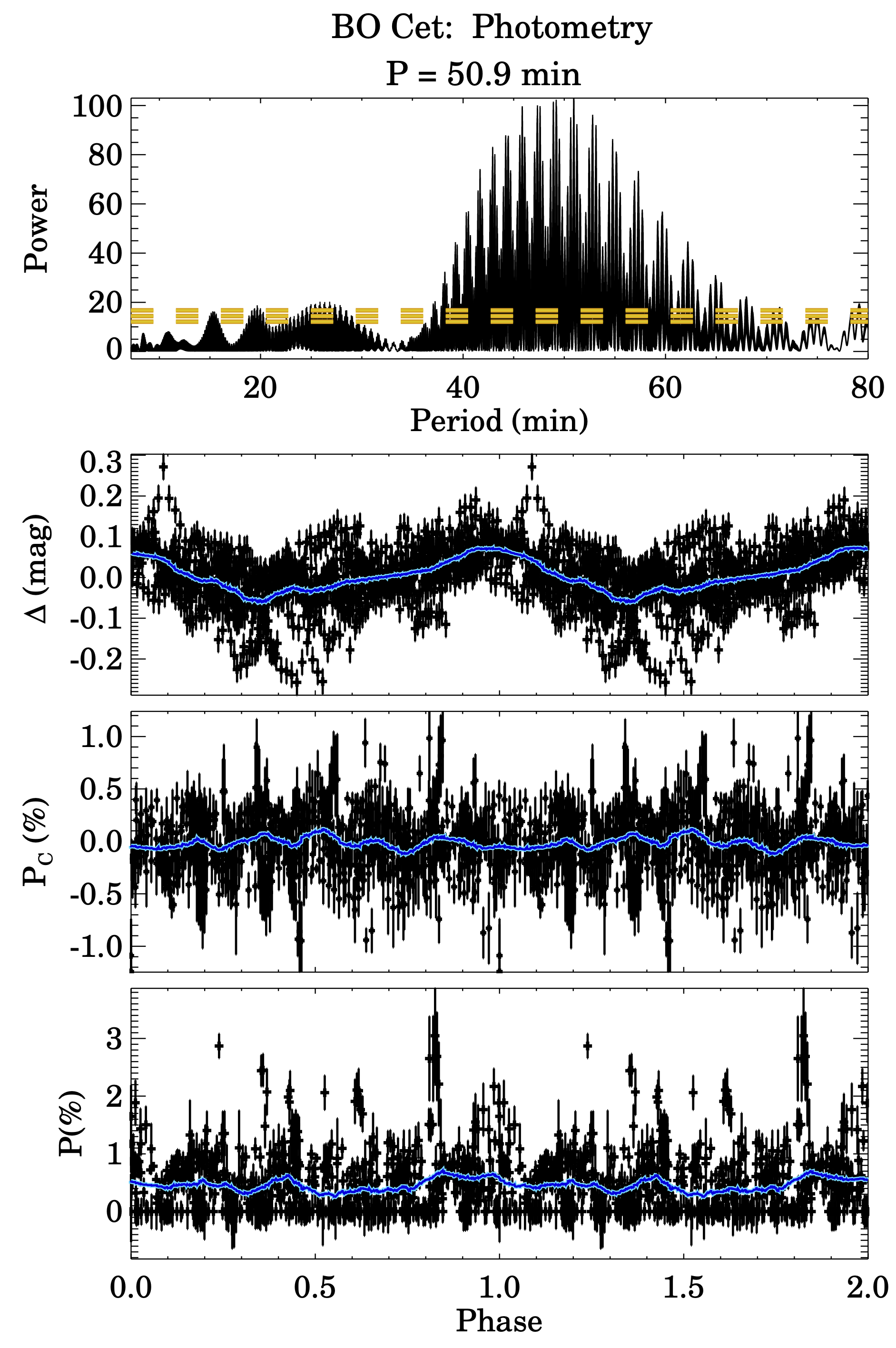}{0.45\textwidth}{}
          \fig{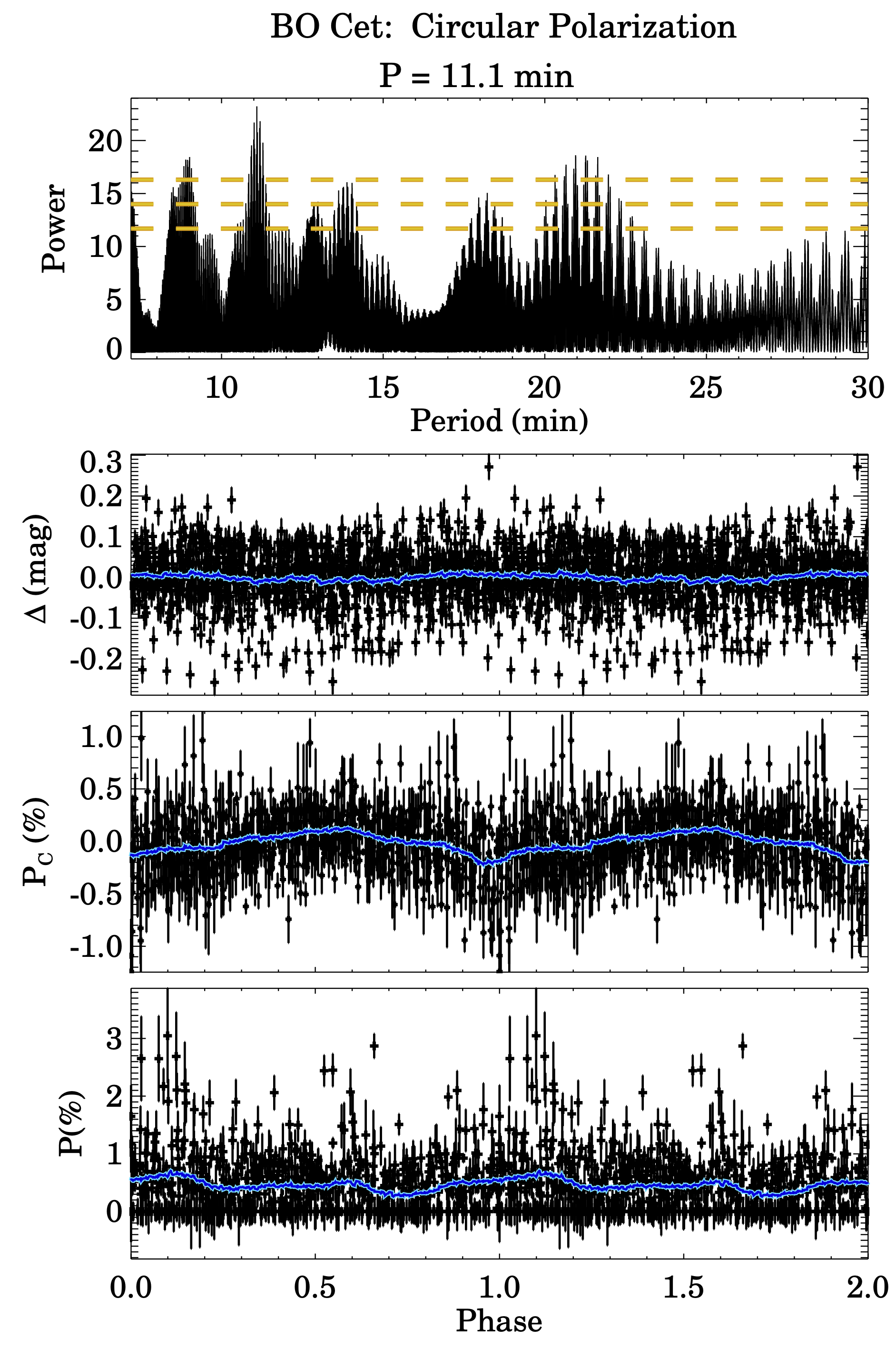}{0.45\textwidth}{}
          }
\caption{Time-series analysis of BO Cet. First column: (Top panel) \replaced{Lomb-Scargle}{LS} periodogram of the photometric data. The second, third, and fourth panels (from top to bottom) exhibit the \replaced{photometry, circular polarimetry (P$_{\rm C}$) and linear polarimetry (P) data}{magnitude, the percentage of circular polazation (P$_{\rm C}$) and the percentage of linear polarization (P) data}, folded with the 50.9~min period and an arbitrary zero phase. \replaced{Second column: same analysis as in the first column}{The second column follows the same structure of  the first one}, but the periodogram \replaced{is calculated for}{corresponds to} the circular polarimetry data, and the curves are folded with the 11.1~min period. The dashed orange lines represent the FAP levels of 1\%, 0.1\%, and 0.01\%, calculated according to \citet{Scargle_1982}. The blue lines \replaced{represent the data averaged by phase}{correspond to a central-moving average considering the points inside an interval of 0.05 in phase.}
\label{FigBO_Cet.1}}
\end{figure*}

The following analysis includes only data from 2010 October 6, 11, 12, and 2016 October 19 (see Table \ref{tab_data}), which do not show significant correlation with the measurements of the field stars, as described in Sec.~\ref{sec_analysis}. The photometric data were pre-whitened at the orbital period before the computation of the \replaced{Lomb-Scargle}{LS} periodograms. The power spectrum peaks at P~=~50.9~$\pm$~\replaced{6.3}{0.5}~min (Fig.~\ref{FigBO_Cet.1}). The strong 1-day aliasing structure hinders the determination of the exact value of this periodicity. But it is most likely real since it is well above the 0.01\% false-alarm probability (FAP) line and is consistently present in the individual nights (not shown here) that are spread over a 6-years range. This peak is also present in the periodogram with no pre-whitening. The magnitude phase diagram shows a clear modulation with a semi-amplitude of around 0.05~mag (Fig.~\ref{FigBO_Cet.1}, second panel from top to bottom).
A harmonic at around 25~min is present in the periodogram. Other peaks are seen at  19.7~min and 15.3~min. 

The (no pre-whitened) circular polarimetric data show a periodic signal at 11.1~$\pm$~\replaced{0.3}{0.08}~min, as shown in the power spectrum of Fig.~\ref{FigBO_Cet.1} (see top panel of the second column). 
This period is not present in the field stars, \replaced{what}{which} gives us some confidence that it is not an artefact. The circular polarization data folded with the 11.1~min period reveals a sinusoidal modulation having both negative and positive polarization excursions and a semi-amplitude of around 0.1\% (see the solid \deleted{the }blue line).

\replaced{Generally, l}{L}inear polarization measurements do not follow a normal probability distribution and are intrinsically biased \citep{Clarke_1983}. In the case of BO~Cet, the bias correction applied to the linear polarization data results in many points of null polarization. Even so, we performed the LS analysis. The power spectrum shows a variety of peaks in the region of 50~min, the same region of the strong signal in the photometry periodogram. The power of those peaks is slightly higher than the level of 0.01\% of FAP. The curve folded using the strongest peak, at 54.4~$\pm$~0.9~min, does not show any coherent variation, therefore we do not show any figure related to the linear polarization data.

The main circular polarization peak at 11.1 min has adjacent structures \replaced{centred}{centered} at about 9~min and 14~min (Table~\ref{table:periods}), which are consistent with the positive and negative beats of the main peak with the 50.9~min photometric main period. The presence of the beat periods between the main photometric and polarimetric peaks supports the reality of these periods. The periodicity of 11.1~min found in circular polarization corroborates that BO~Cet could harbour magnetic accretion, as suggested by \citet{rodriguez/2007a}. In this \replaced{intermediate polar}{IP} scenario, the photometric signal at $\sim$51~min could be associated with a continuum radiation source located in the inner disk region.  
In \replaced{a magnetic accretion scenario}{particular}, this region could be at the magnetosphere radius, where the mass flow starts to follow the magnetic field lines and leaves the orbital plane.

\citet{rodriguez/2007a} found a period of 19.9~min from the radial velocities of the  H$\alpha$ line wings, consistent with a 19.6~min period from the equivalent widths of the H$\alpha$ blue wing.  They suggested this to be associated \replaced{to}{with} the WD rotation period. As the origin of the circular polarization is more directly connected with the WD rotation, we suggest that the WD spin period is 11.1~min. Interestingly, our photometric power spectrum also shows a peak at around 19.7~min, which could be related to the line-emission source detected by \citet{rodriguez/2007a} diluted by continuum emission.

\subsection{SW~Sex} \label{results_swsex}

SW~Sex, the prototype of the class, was discovered in the Palomar-Green survey by \cite{Green_1982}. Follow-up spectroscopy and photometry  \citep{Penning_1984} showed high-excitation emission lines, a deep eclipse of 1.9~mag, and an orbital period of 3.24~h. A refined orbital period, 0.1349384411(10)~d, was obtained by \citet{Groot_2001}.
\replaced{SW~Sex has magnitudes ranging from $\sim$14.4 (high state) to $\sim$16.1 (low state) in the V band, according to observations available in the American Association of Variable Star Observers website}{SW~Sex has been observed in two brightness states considering its magnitude out of eclipse, which has mean values of $\sim$14.2 (high state) and $\sim$15.1 (low state) in 
unfiltered observations calibrated with Johnson V band zeropoint, according to observations available in the American Association of Variable Star Observers website} (AAVSO). Spectroscopy performed in the bright state showed the typical central absorption in the Balmer emission lines between phases 0.4 and 0.6 \replaced{\citep{szkody/1990}}{\citep[e.g.][]{szkody/1990}}. Spectrophotometric observations during the faint state of the system did not exhibit the absorption feature and emission-line flaring was not detected \citep{dhillon/2013}. No periodicity \added{ other than that due to the eclipses} in broad band photometry is reported for this system. Circular polarimetr\replaced{y}{ic} observations by \cite{Stockman_1992} revealed no significant circular polarization in the 3200~--~8600~\AA~ range, with values as low as 0.05~$\pm$~0.14\% in a single 8~min integration and -0.03~$\pm$~0.05\% in 16~min. However, in a paper about the detection of circular polarization in LS~Peg, \cite{rodriguez/2001} made a side note about a period of 28~min detected in the B band circular polarization of SW~Sex.

\begin{figure*}
\gridline{\fig{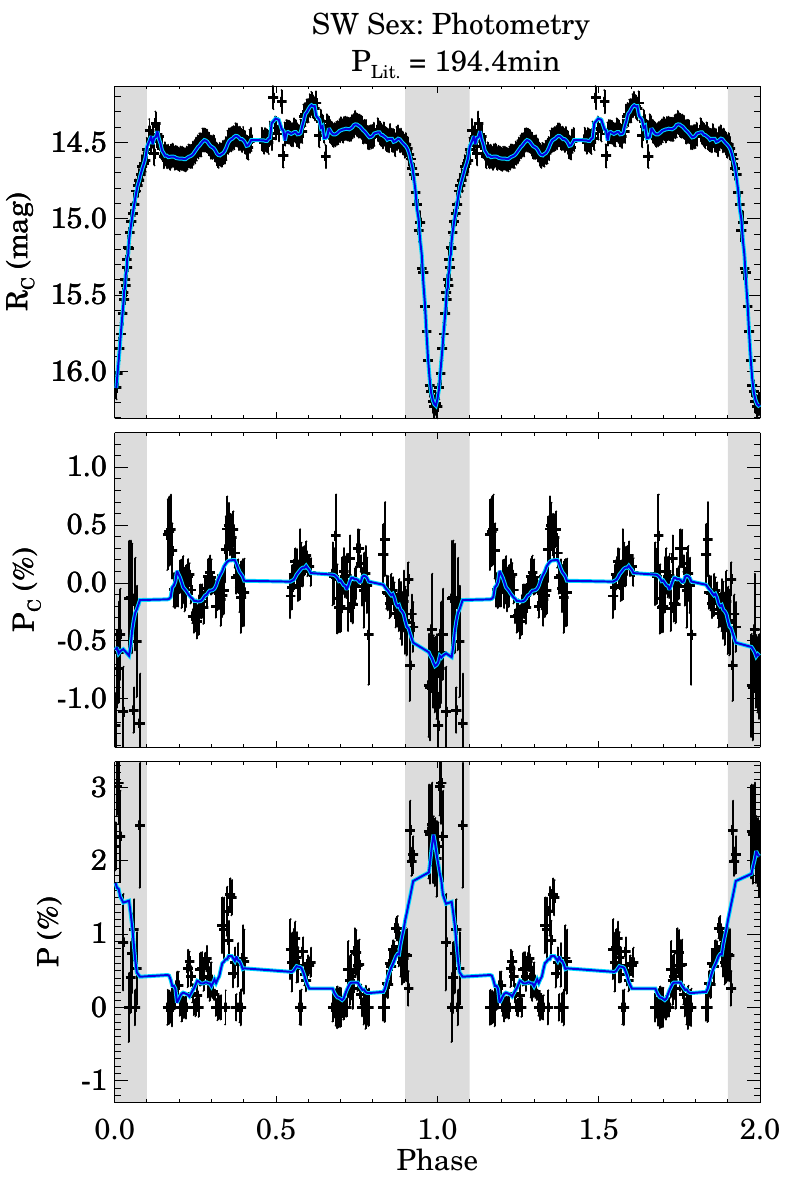}{0.32\textwidth}{}
          \fig{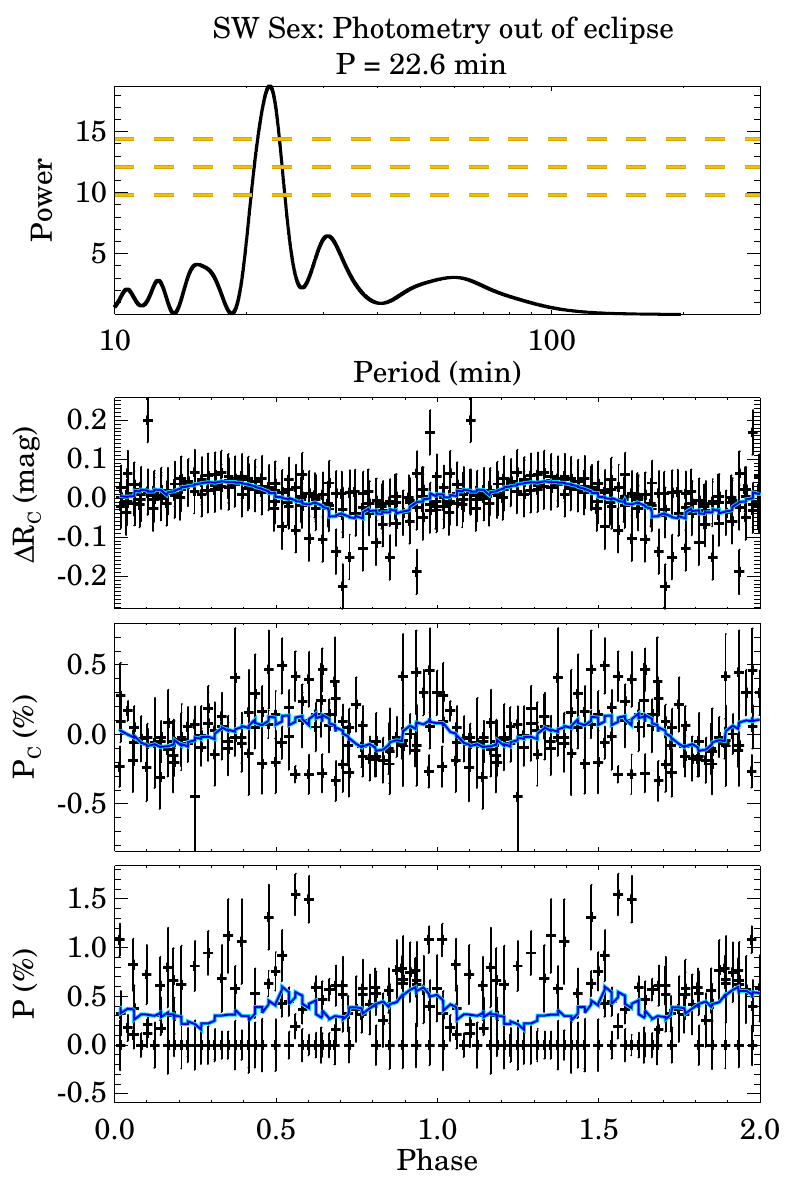}{0.32\textwidth}{}
          \fig{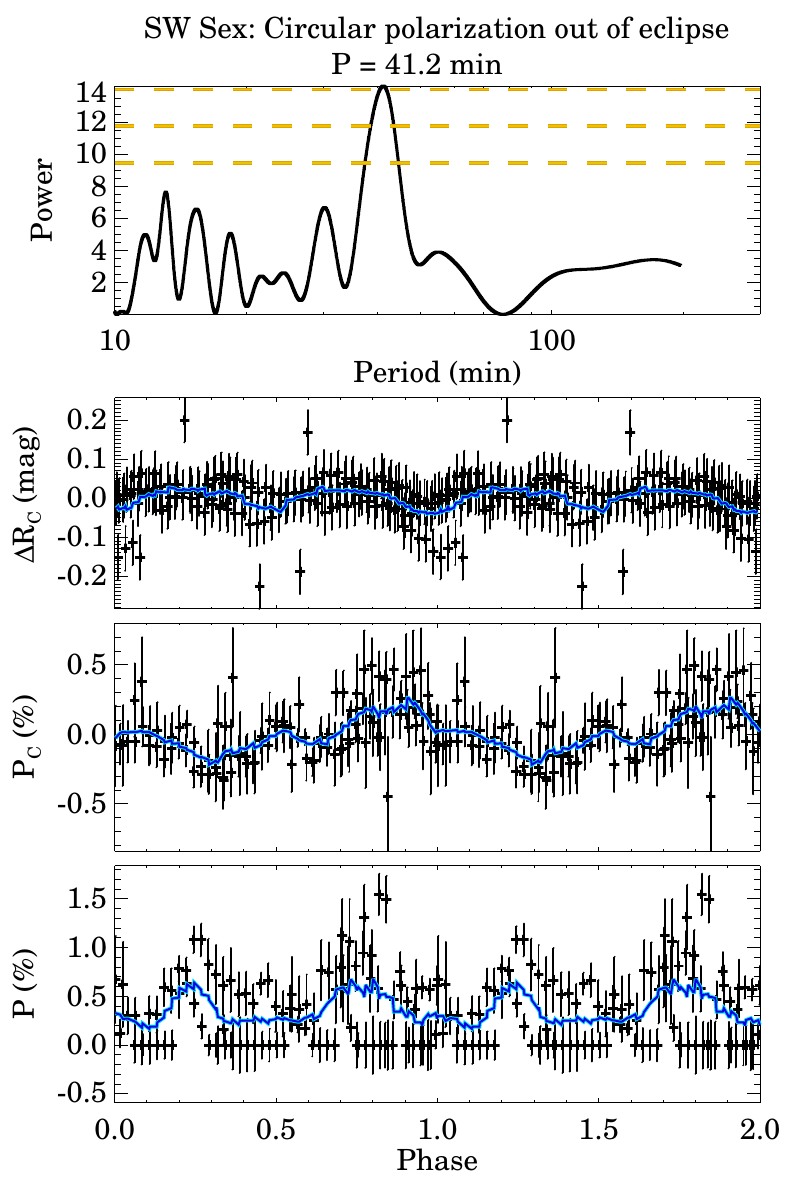}{0.32\textwidth}{}
          }
\caption{Time-series analysis of SW Sex. First column: Phase diagrams of the photometry, circular polarimetry (P$_{\rm C}$) and linear polarimetry (P) data, folded with the orbital period and T$_{0}$ = 2456746.704430 HJD. The eclipse phases are highlighted by the gray bar.  The points in this region were not considered in the analysis presented in the following columns. Second column: the top panel shows the \replaced{Lomb-Scargle}{LS} periodogram of the photometry, after subtracting the orbital modulation (see text). The panel below presents the phase diagram of the magnitude folded with P~=~22.6~min and arbitrary zero phase. The third and \replaced{forth}{fourth} panels from top to bottom show the folded circular and linear polarization, respectively. The last column follows the same structure of the second column, but \replaced{fo}{for} circular polarization. The lines are colour-coded as in Figure~\ref{FigBO_Cet.1}.
\label{fig_SWSex.1}}
\end{figure*}

SW~Sex was observed by us \replaced{in}{on} one single night \replaced{on March 29, 2014}{(March 29, 2014)}. A time series of R$_{\rm C}$ measurements was made during 3.9~h covering one complete orbital cycle and one total eclipse. The mean magnitude out of eclipse in our photometric data is R$_{\rm C}$=~14.66~mag, indicating that the system was in a high brightness state. The folded light curve shows a deep eclipse with amplitude of 1.6~mag, as expected (Fig.~\ref{fig_SWSex.1}, first column). 

The folded polarization data reveal an increase in the absolute values of circular and linear polarizations during the eclipse (Fig.~\ref{fig_SWSex.1},  bottom panels). This could be interpreted as a polarized emission component that is diluted by an unpolarized light source \replaced{that is (partially) occulted during the eclipse}{out of eclipse. During the eclipse, part of the unpolarized component is occulted}, thus reducing the dilution and making the net polarization of the system larger. Reasonable guesses as to the origin of the polarized and unpolarized components are the \replaced{post-shock}{PSR} region of a magnetic accretion column and the accretion disk, respectively. The eclipse of the  \replaced{post-shock}{PSR} region could also occur. It would last less than the disk eclipse, because the \replaced{post-shock}{PSR} region is much smaller. In the polars FL~Cet and
MLS110213~J022733+130617, the \replaced{post-shock}{PSR} region is eclipsed during an interval of about 0.05 in \replaced{phase}{the orbital cycle} \citep[][respectively]{flcet, Silva2015}. In this case, the polarization should drop to zero around the middle of the eclipse, producing an ``M" shaped polarization pattern. As it is not observed, we can conclude that the \replaced{post-shock}{PSR} region is not eclipsed in SW~Sex. 

To perform the period search, we removed the points associated with the eclipse from the dataset and then subtracted the orbital modulation using a third-order polynomial fit to the photometric data  \citep[e.g.][]{Basri_2011, Ferreira_Lopes_2015}. The resulting power spectrum has one significant peak at 22.6~$\pm$~\replaced{3.8}{1.4}~min (see Fig.~\ref{fig_SWSex.1}). 
This peak is also present, but with a lower power, if we do not \replaced{filter the data by}{subtract} the orbital modulation. The light curve folded at this period displays a modulation with an amplitude of about 0.11~mag. 

The period search in the polarimetric data was also performed removing the eclipse points. 
The periodogram of circular polarization shows a peak at 41.2~$\pm$~\replaced{7.7}{8.5}~min. The mean polarization in the phase diagram ranges between -0.2\% \replaced{to}{and} 0.2\% (blue line in circular polarization panel of the last column of Fig.~\ref{fig_SWSex.1}). Interestingly, the photometric period (22.6~min) is consistent with the first harmonic of the period found in polarimetry (41.2~min), considering the uncertainties in the peak position\added{s}.
The phase diagrams of the photometry using these two periods in Fig.~\ref{fig_SWSex.1} illustrate this. The linear polarization folded with the 41.2~min period shows a relatively well organized phase diagram. However, the highest peak in linear polarization occurs at 12.7~$\pm$~0.7~min (below the adopted FAP levels) and the corresponding phase diagram is very noisy. Hence, we do not discuss it here. Probably, the time-series analysis of the linear polarimetry is affected by the large number of measurements consistent with zero, due to the large errorbars. 

Some IPs \citep[e.g. V405~Aur and UU~Col,][respectively]{Piirola_2008, Katajainen_2010ApJ} show two maxima in the flux phase diagram, associated with positive and negative circular polarization.
Our data suggest that this can also be true for SW~Sex. This finding and the increase in the polarization during the eclipse support the presence of a magnetic WD rotating at around 41~min in SW~Sex. Additional polarimetric observations having a longer time span would be useful to confirm these results.

\subsection{V442~Oph} \label{results_v442oph}

V442~Oph was identified as a CV by \cite{Szkody_1980}. 
\replaced{Its magnitude varies in the range V~=~12.6~--~14.0~mag \citep{Ballouz_2009}.}
{The typical V band brightness of V442~Oph fluctuates \replaced{around between}{between around} 13.2 and 14.2~mag \citep{Szkody_1980,Szkody_1983,Patterson_2002}. However, AAVSO data show that it can \added{be} as faint\deleted{er} as 15.5~mag. This is somewhat confirmed by \citet{Patterson_2002}, who affirmed that the 14 mag state corresponds to a high state. However, those authors do not mention the maximum magnitude of V442 Oph in their data.}
Some spectral features point to a classification as a low-inclination SW~Sex star \citep{Hoard_2000}: single-peaked emission lines, strong \heii emission line, a transient absorption at 0.5 phase in the Balmer and \replaced{He~I}{\ion{He}{1}} lines, and a high-velocity component in H$\alpha$ at~$\approx\pm$1900~km~s$^{-1}$. Their spectroscopic dataset settled the orbital period as 0.12433~d (2.98~h). Long photometric campaigns show persistent modulations with periods at 4.37(15)~d and 0.12090(8)~d, which are interpreted as the wobble period of the disk and a negative superhump, respectively \citep{Patterson_2002}. At higher frequencies, the system displays rapid flickering and periodicities around 1000~s, interpreted as QPOs. \replaced{Over}{In}  the QPO structure \replaced{in}{of} the periodogram of the best quality data, there are two distinguishable peaks at 74.0 and 89.9 cycles per day (19.5 and 16.0\,min). Circular polarimetry obtained in two nights provided 0.01\%~$\pm$~0.05\% and -0.03\%~$\pm$~0.08\% with 16~min and 8~min of total integration times, respectively \citep{Stockman_1992}.
These values are consistent with no net circular polarization averaged over the time interval of the measurements.

We observed V442~Oph for a total of 4.06~h spread over two nights: 2014, March 29 and 2014, July 20 (Table~\ref{tab_data}). Light curves in R$_{\rm C}$ and V filters were obtained using exposure times of 20~s and 30~s, respectively. 
The time series has some gaps repeating on time scales of half an hour, so we limit our analysis to a maximum period of 30\,min.
The mean magnitudes during each day of our observations are quite similar, R$_{\rm C}$~=~13.37~mag and V~=~13.60~mag. Such values indicate that we observed V442~Oph \replaced{near its low state}{in its typical high state}. 

\begin{figure*}
\gridline{\fig{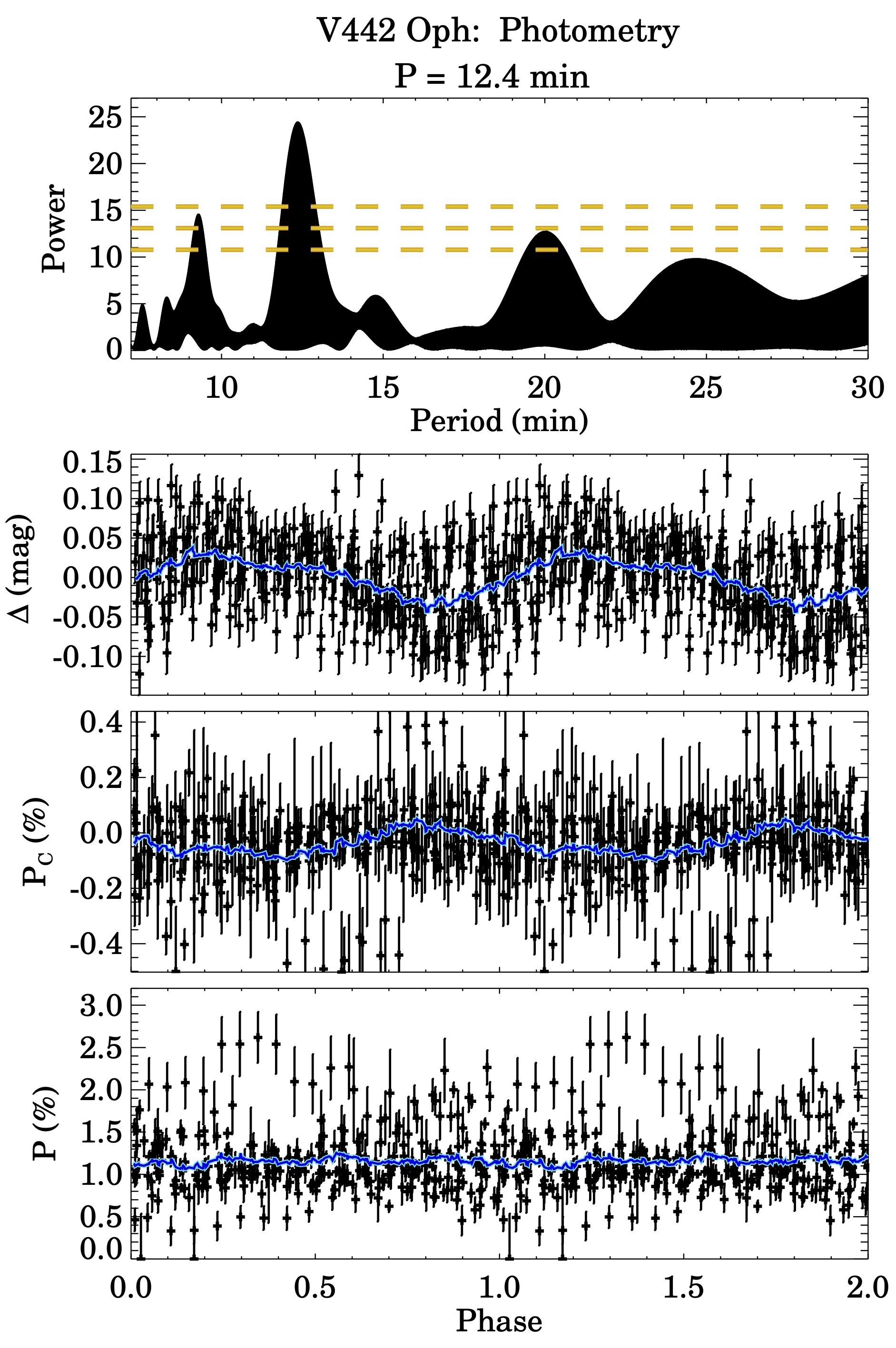}{0.33\textwidth}{}
          \fig{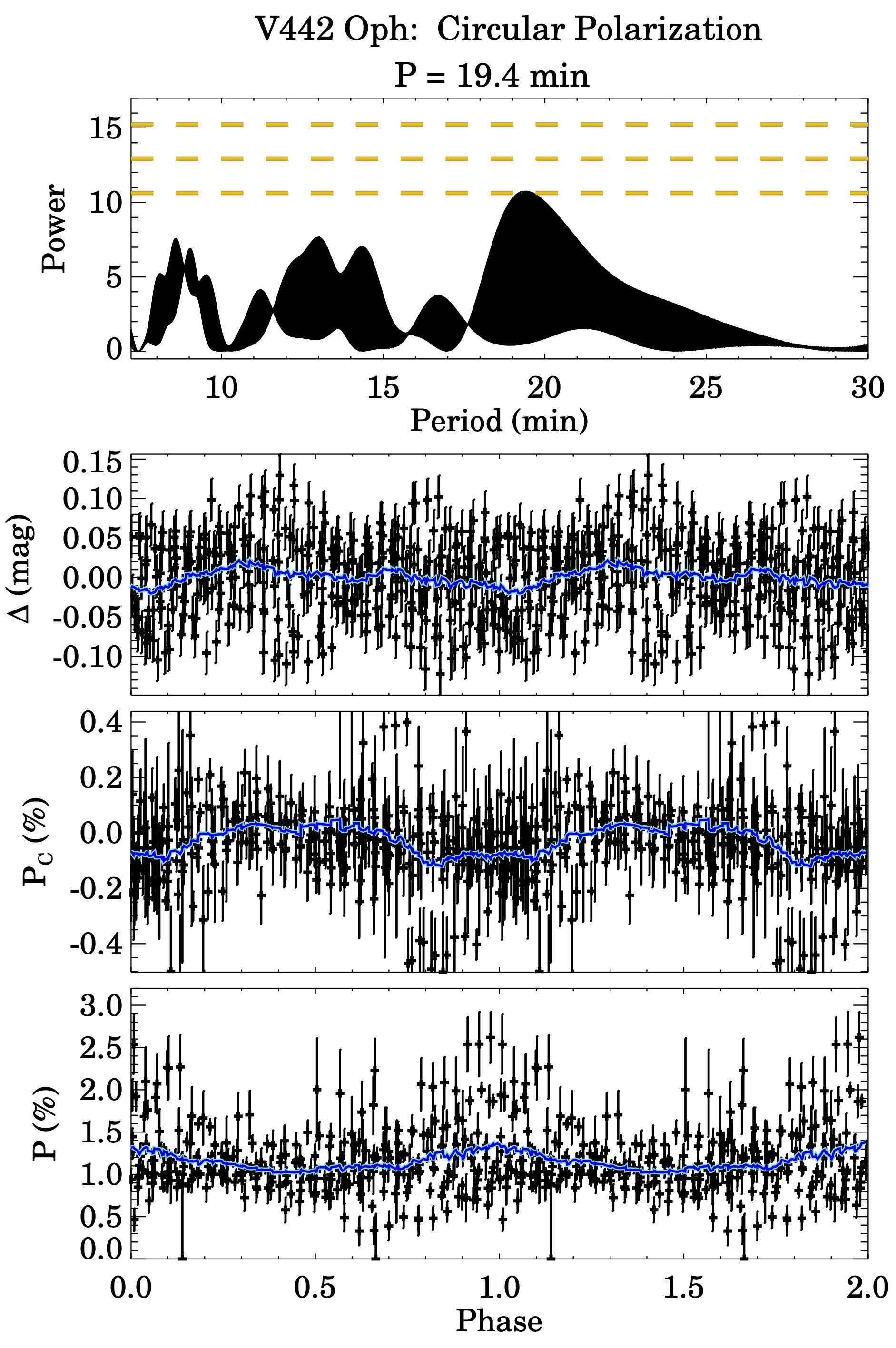}{0.33\textwidth}{}
          \fig{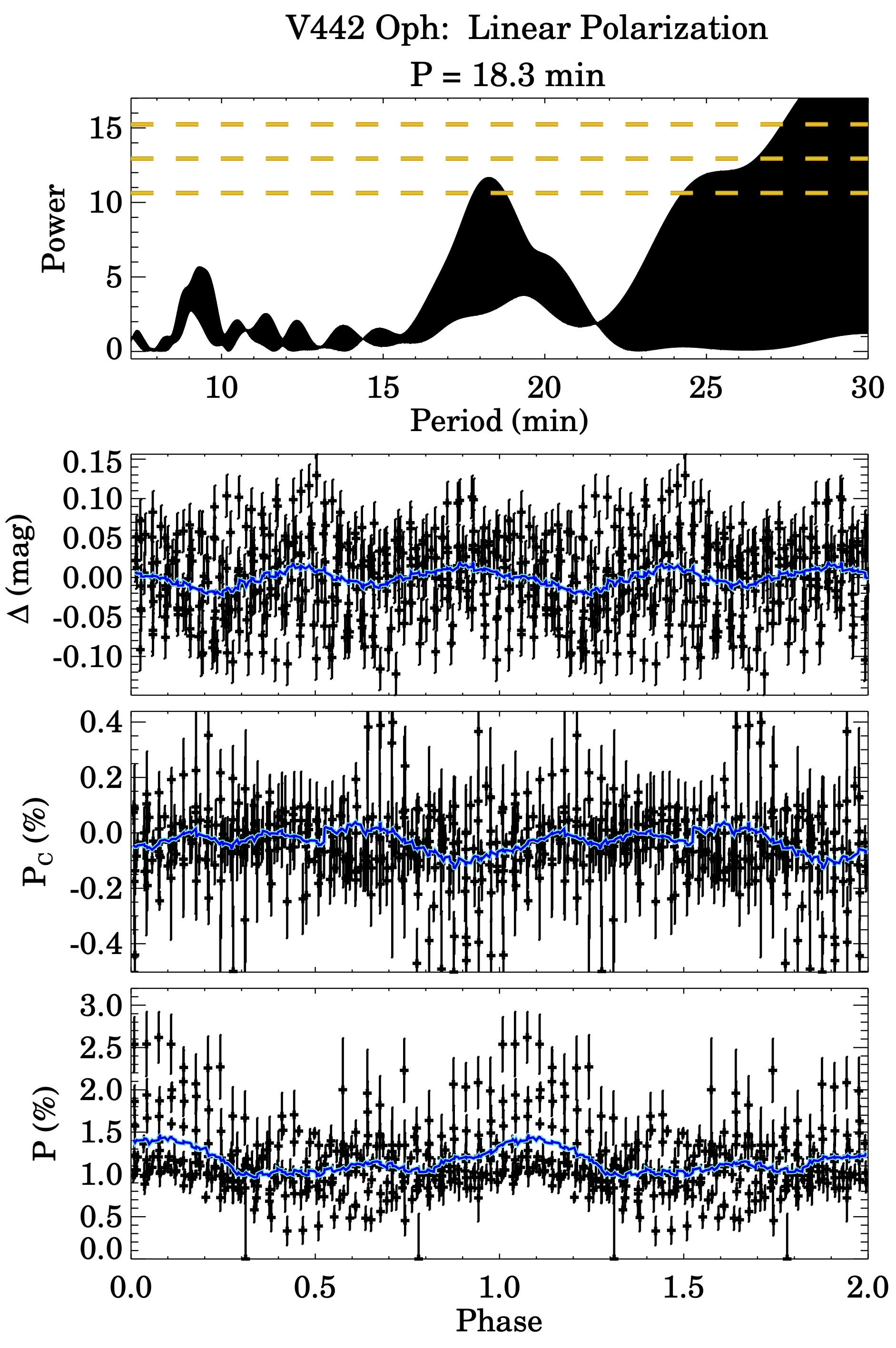}{0.33\textwidth}{}
          }
\caption{Time-series analysis of V442~Oph. First column, \added{from} top to bottom: the \replaced{Lomb-Scargle}{LS} periodogram from photometry; the phase diagram of the magnitude folded with 12.4~min and an arbitrary zero phase; the same as above for the circular polarization; and the phase diagram of the linear polarization using the same period. The second and third columns follow the same structure of the first column, but the periodograms are done using the circular and linear polarization data, respectively. The orange and blue lines follow the definition of previous figures.
\label{FigV442ph.1}}
\end{figure*}

We performed a period search in \added{the} R$_{\rm C}$ and V photometric time series independently and found a period of around 12~min in both cases. No pre-whitening or detrending was applied to the data.
Finding the same period in two independent data sets, obtained about four months apart, suggests the presence of  a stable periodicity in V442~Oph. 
The power spectrum of the combined photometry has a prominent peak located at 12.4~$\pm$\replaced{0.5}{0.09}~min (see first column of Fig.~\ref{FigV442ph.1}).
The data folded using this period and an arbitrary epoch at HJD~2456859.4033 show a near sinusoidal profile with a semi-amplitude around 0.04~mag (second panel). The folded circular polarimetric curve also shows a slight modulation with an amplitude of around 0.1\% (blue line in the third panel). A peak at 19.9\,min with 0.1\% false-alarm probability merits attention, since it also appears in the circular and linear power spectra --- see below.

The power spectra of the circular and linear polarization show peaks at 19.4\,$\pm$\,\replaced{1.7}{0.4}\deleted{\,min} and 18.3\,$\pm$\,1.1\,min, \added{respectively,} which are below the FAP lines of 0.1\% (Fig.~\ref{FigV442ph.1}, middle and right columns). The data folded at these periods do not show clear modulations, but we suggest that there could be a real periodicity of the system around 19\,min. Besides being present in the periodograms of flux, circular and linear polarizations - even \replaced{if}{though} with a weak signal - a similar period is also found by \citet[see Table \ref{table:periods}]{Patterson_2002}. They commented that the modulation  semi-amplitude at this frequency is around
0.01~mag and that it is not easily detectable among other variability also present in the light curves, features which are consistent with our folded curves. As this period is present in \added{our} polarimetric data, its natural explanation is the WD rotation.
\deleted{Our observations were obtained during a low state of V442~Oph, which can favour the detection of the WD spin period, as discussed by \citet{Patterson_2002}. They argued that the spin period of the WD in AE~Aqr (33~s) is seen as a stable modulation only in the low state: in the high state, the flux modulation has a QPO structure at around 35~s.}

In spite of the low signal, we suggest that the 19.4\,min modulation found in V442~Oph is due to the WD rotation. However, we have no explanation for the 12\,min modulation seen in the total flux. In particular, we could not explain this modulation as a beat between other periodicities of the system.

\subsection{V380~Oph} \label{results_v4380oph}

V380~Oph was reported as a nova-like CV with orbital period of 3.8~h based on spectroscopic observations \citep{Shafter_1985}. Using archival data and new photometric observations, \citet{Shugarov_2005} showed that \vto is usually at 14.5~mag, but underwent a faint state (around 17~mag) in 1979. Hence, they suggested that the system can be classified as a VY~Scl object.\deleted{This type of \replaced{novalike}{nova-like variable} shows episodic reductions in brightness, so they were originally termed as anti-dwarf nova. They do not show outbursts.} Indeed, the VY~Scl behaviour of V380~Oph is confirmed by additional data presented by \citet{Shugarov2016} and the light curves available at AAVSO. 
\cite{Shugarov_2005} also found two photometric periods\replaced{:}{,} 3.56~h and 4.51~d, interpreted as due to negative superhumps and an eccentric wobbling accretion disk, respectively. Time-resolved spectra obtained by \citet{rodriguez/2007a} show the standard characteristics of the SW~Sex class, single-peaked emission lines and a high-velocity S-wave in the H$\alpha$ trailed spectra.
Their radial-velocity curve \replaced{has a}{provided an orbital} period of 3.69857(2)~h, consistent with the period obtained by \cite{Shafter_1985}\deleted{, but confirming and improving the orbital period}. 
They also suggest the presence of rapid flaring in the line wings with a periodicity of 46.7~$\pm$~0.1~min, which they proposed as the rotation of a magnetic \replaced{white dwarf}{WD}. Optical photometry of \vto performed by \citet{Bruch2017} is consistent with an orbital modulation and reveals \replaced{important}{a} flickering activity\deleted{, behavior} that is compatible with \added{those of} VY~Scl systems. White-light circular polarimetry was performed by \citet{Stockman_1992}, who obtained a signal consistent with zero, P$_{\rm C}$~=~-0.12~$\pm$~0.16\%, in a \replaced{total}{short} 8 minute integration.

\begin{figure*}
\gridline{\fig{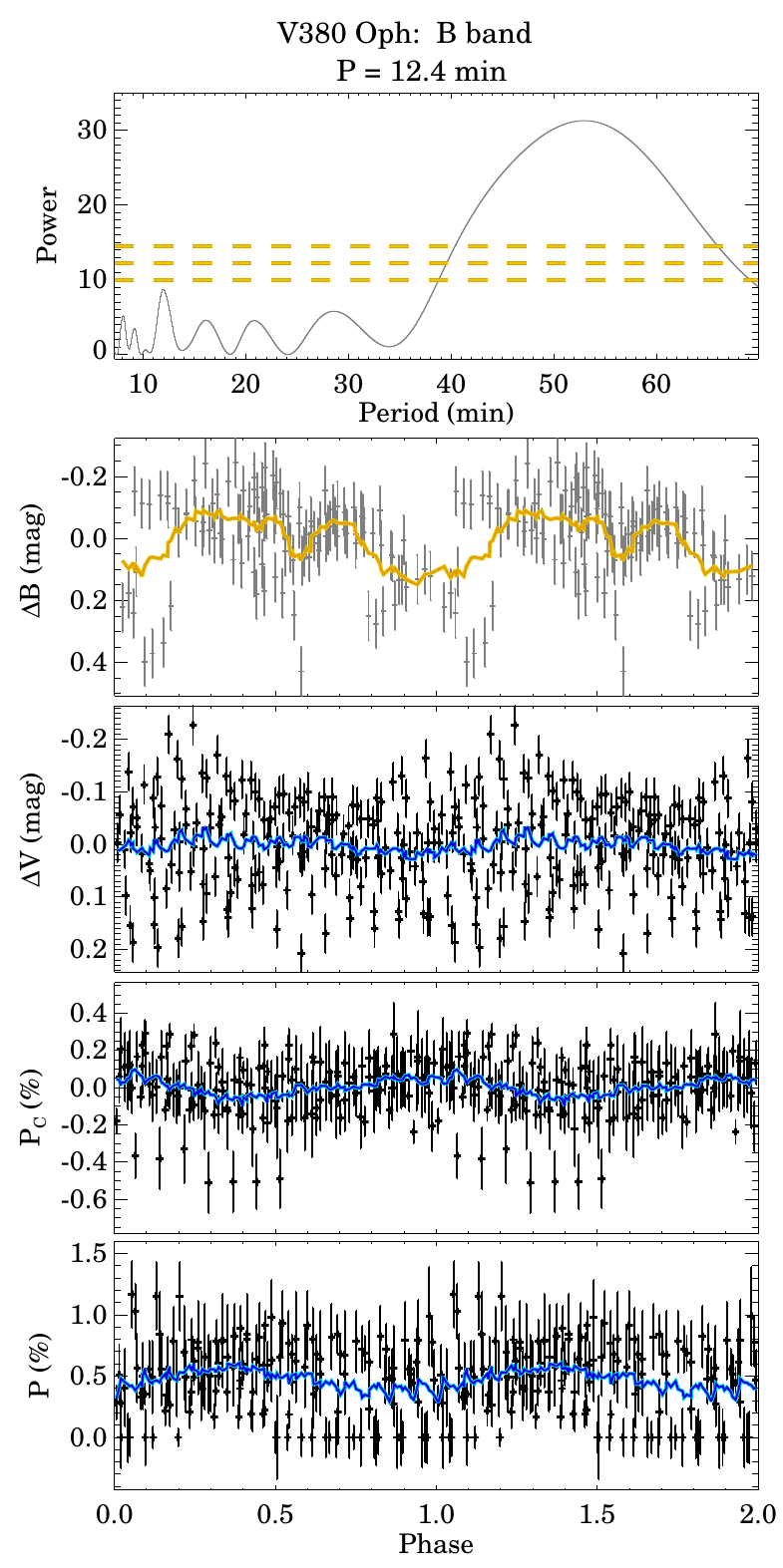}{0.32\textwidth}{}
          \fig{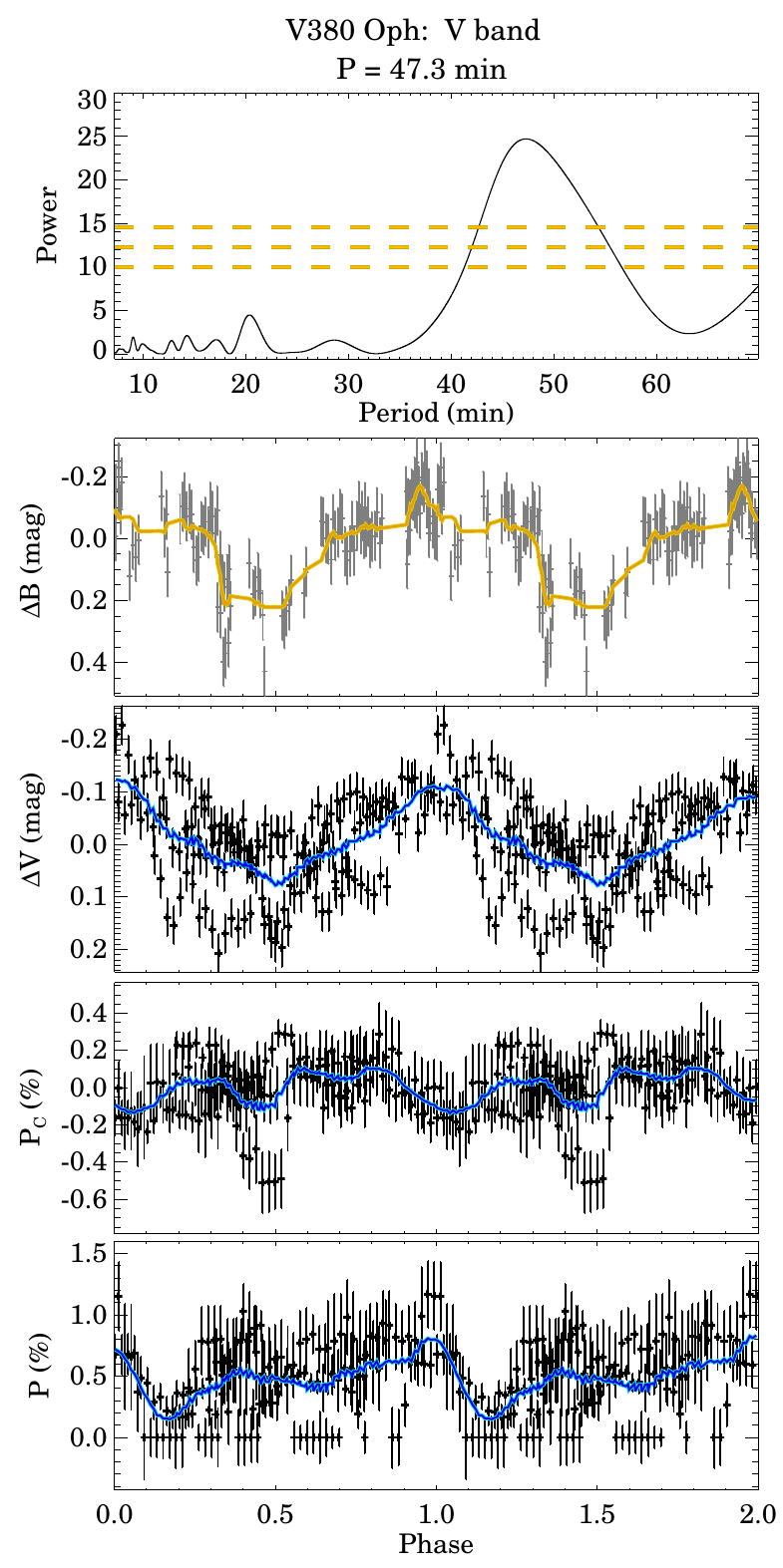}{0.32\textwidth}{}
          \fig{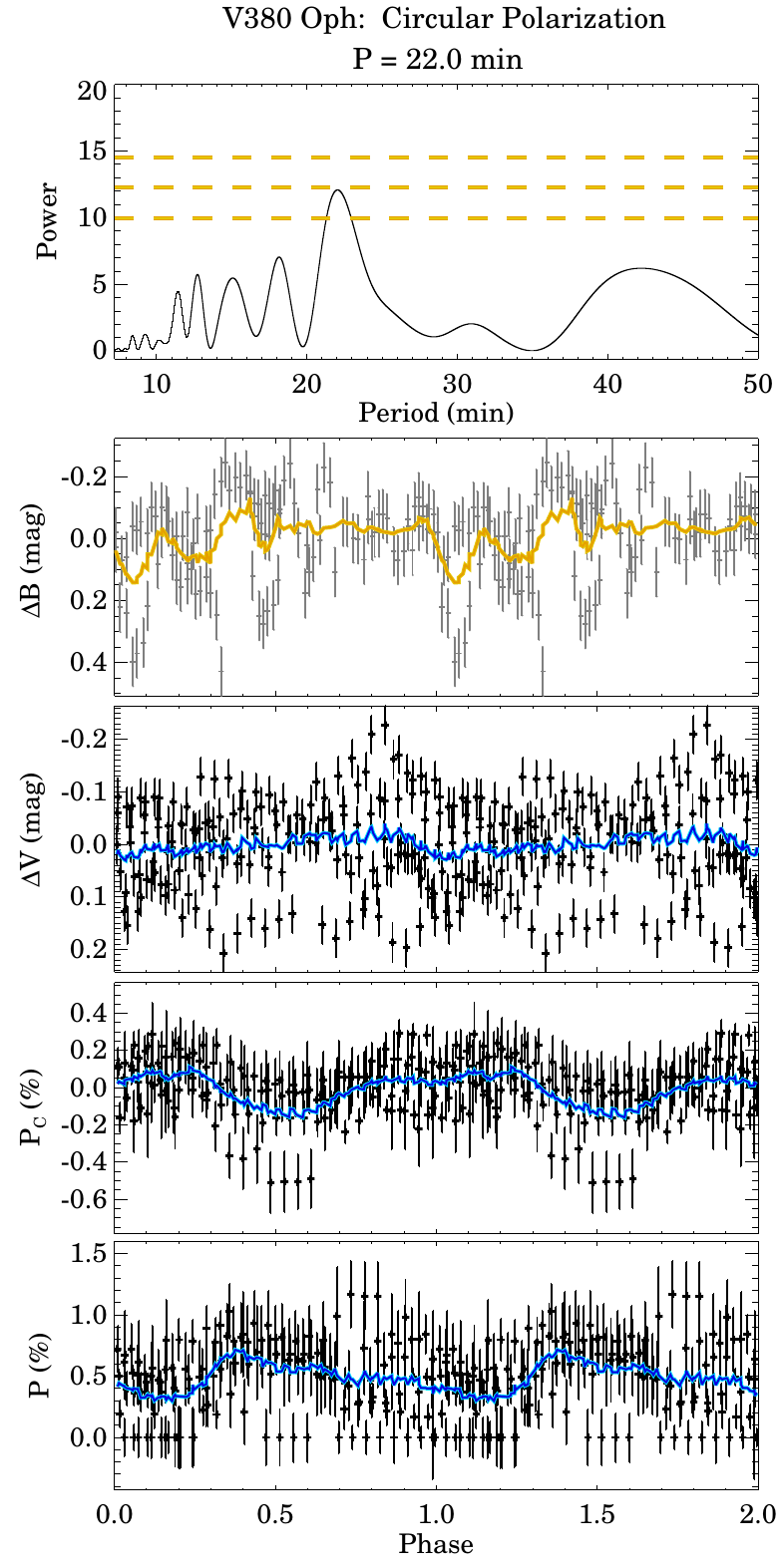}{0.32\textwidth}{}
          }
\caption{Time-series analysis of V380~Oph.  First column: B band photometric data from SOS/SAS observatories. Second column: V band photometric data from OPD observatory. Third column: V band circular polarimetric data from OPD observatory. From top to bottom: the power spectrum; the phase diagrams of B band and V band photometry, and circular and linear polarizations. The coloured lines are as defined in Figure~\ref{FigBO_Cet.1}.
\label{FigV380Oph.1}}
\end{figure*}

V380~Oph is the only object of our sample for which we have additional data obtained from observatories other than OPD (see Section~\ref{sec_observation} and Table \ref{tab_data}). 
We reanalyzed the SOS/SAS V380 Oph photometric data  \citep{Shugarov2016} obtained  from 2002 to 2016 searching for short periods possibly associated with the WD rotation. 
Typically, V380~Oph has a mean magnitude of approximately 15~mag (high state). During three nights on 2015 July 15 to 22, a low state was detected in the R$_{\rm C}$ band with an average brightness of 18.8~mag. 

Initially, we present the periodicity analysis of the time series obtained at SOS/SAS observatories. The V band data were obtained over many years and are \replaced{characterised}{characterized} by data blocks with time spans from around 1 to 4~h. 
The \replaced{Lomb-Scargle}{LS} power spectrum (not shown here) is very complex. It shows a clear signal at around 4.5\,d, as previously reported. The peak is broad and highly structured. A stronger and narrower peak at around 1.3\,d is also present. We have filtered the time series using the main peaks relative to periods on a timescale of days. The periodograms remain very complex with many peaks on timescales of hours and days. No clear signal on a minutes timescale could be found. We also studied the data sets on a daily basis and no clear signal appears. We applied the {\sc clean} technique \citep{clean} in order to verify if we could obtain a better understanding of the photometric variability, but the results are essentially the same as those obtained using the \replaced{Lomb-Scargle}{LS} method. 
The R$_{\rm C}$ data in the bright state are sparser and have shorter daily data blocks compared to the V band. The results are essentially the same for both bands. The R$_{\rm C}$ data set in the low state is composed of 3 daily blocks: no clear periodicity is found. 
The $B$ band periodogram has a very broad feature at around 55~min, which is badly constrained by the short time span of about 2~h. 
Similar features are present in some power spectra of \added{the} V and R$_{\rm C}$ daily data blocks. A peak at 12.4\added{~$\pm$~0.66}~min is present below a FAP of 1\%.  Fig.~\ref{FigV380Oph.1} (left column) shows the power spectrum and the \added{B band} data folded on 12.4~min. There seems to be a coherent modulation of the circular polarization at this period. After \replaced{filtering the data by the orbital period}{removing the orbital modulation}, only the 12~min peak remains, yet below the 1\% FAP.

Our OPD observations were performed in the V band \replaced{in}{on} a single night, 2014 July 19, having a total time span of $\sim$2.7~h with 40~s of exposure time. The system was in the usual high state at around 14.5~mag. The power spectrum of the photometry shows a periodicity at 47.4~$\pm$~\replaced{4.3}{4.9}~min (Fig.~\ref{FigV380Oph.1}, middle column). This period is in the upper limit of the range of reliable periods due to the total time span of the observation. However, it is consistent with the $\sim$47~min flaring period reported by \cite{rodriguez/2007a}.

The circular polarimetry periodogram shows a peak at 22.0~$\pm$~\replaced{2.2}{1.2}~min with a level similar to the 0.1\% FAP (Fig.~\ref{FigV380Oph.1}, right column). After \replaced{filtering the data by the orbital period}{removing the orbital modulation}, we obtained a period at 12.7~$\pm$~\replaced{0.6}{0.4}~min above the 0.01\% FAP (not shown), which is similar to that found in the B band data obtained at the SOS/SAS observatories. In this case, the modulation in circular polarization is similar to that shown in Fig.~\ref{FigV380Oph.1}, left column, therefore we do not present a figure with exactly this period. In the linear polarimetry data, we found 17.7~$\pm$~1.5~min. However, the same period is also found in one field star, consequently this period can be related to instrumental effects and is suspicious. 

In addition to its complex photometric behaviour, V380~Oph also shows variation in the measured values of the radial velocity amplitude depending on its brightness state \citep[see][and references therein]{Szkody_2018}, which could be related to the variation of the accretion-disk size. \citet{Zellem2009} obtained the UV spectrum of \vvto, which is too red to be fitted with an optically-thick accretion-disk model. A possible interpretation is an accretion disk truncated in its inner parts in an IP-like configuration. These facts together with the presence of line flaring (Table~\ref{table:periods}) and a possible periodicity in circular polarimetry may indicate magnetic accretion. But, we consider additional polarimetric and spectroscopic observations are important to corroborate these findings.

\subsection{LS~Peg}  \label{sec_results_lspeg} 

LS Peg was classified as a CV by \citet{Downes_1988}. It has high and low states ranging from V~=~12 to 14~mag \citep{Garnavich_1992}. UV spectra at the high state led \cite{Szkody_1997} to propose LS~Peg as a non-eclipsing SW~Sex object.
\deleted{Later, \citet{Szkody_2018} suggested that LS~Peg may be a magnetic CV with a WD rotation period around 20~min due to a conspicuous photometric modulation. However, it is uncertain if this period is stable in time. Examples of periods cited in the literature are: 19~min, 16.5~$\pm$~2~min, and 20.7~$\pm$~0.3~min \citep{Garnavich_1988, Garnavich_1992, Szkody_1994p, Taylor_1999, Szkody_2001}.} 
\cite{Taylor_1999} estimate an orbital period of 0.174774(3)~d based on radial velocity curves. 
\added{The same orbital period was almost simultaneously reported by \citet{Martinez_1999}.}
The first detection of modulated circular polarization in an SW~Sex system was obtained for LS~Peg: the \deleted{``broad-band"} circular polarization values were the result of the integration in the 3900~--~5070 \AA~range of spectropolarimetric data and are consistent with a periodicity of 29.6~$\pm$~1.8~min and amplitude of $\sim 0.3 \%$ \citep{rodriguez/2001}. Those authors suggested that this modulation corresponds to the WD spin. That paper also shows the presence of flaring in the H$\beta$ emission line with a period of 33.5~$\pm$~2.2~min, which the authors interpreted as the beat period between the WD spin period (29.6\,min) and the orbital period. \cite{baskill/2006} reported a detection of a period at 30.9~$\pm$~0.3~min in X-ray data obtained by ASCA~SISO (2~--~8 keV). However, this modulation was not confirmed by the XMM-Newton observations of LS~Peg in the 0.1~--~12~keV energy range \citep{Ramsay_2008}. These authors \replaced{claim}{mentioned} that the X-ray spectrum of LS~Peg is similar to those of \replaced{intermediate polars}{IPs}.
\added{Later, \citet{Szkody_2018} suggested that LS~Peg may be a magnetic CV with a WD rotation period around 20~min due to a conspicuous photometric modulation. However, it is uncertain if this period is stable in time. Examples of periods cited in the literature are: 19~min, 16.5~$\pm$~2~min, and 20.7~$\pm$~0.3~min \citep{Garnavich_1988, Garnavich_1992, Szkody_1994p, Taylor_1999, Szkody_2001}.} 

Our data on LS~Peg were obtained on 7 nights between 2010 and 2019\deleted{,} in \added{the} V and R$_{\rm C}$ bands (Table~\ref{tab_data}). \replaced{P}{The p}hotometric light curves show a mean magnitude of 11.9~mag and 11.7~mag in \added{the} V and R$_{\rm C}$ bands, respectively. Data from two nights were removed from the periodicity analysis (see Sect.~\ref{sec_analysis}) that was performed \replaced{using the light curve in all observed bands, the light curves combined by filters, and the individual nights.}{combining the data in three ways: all observed bands, combining data by filters, and the individual nights.} 

\begin{figure*}
\gridline{\fig{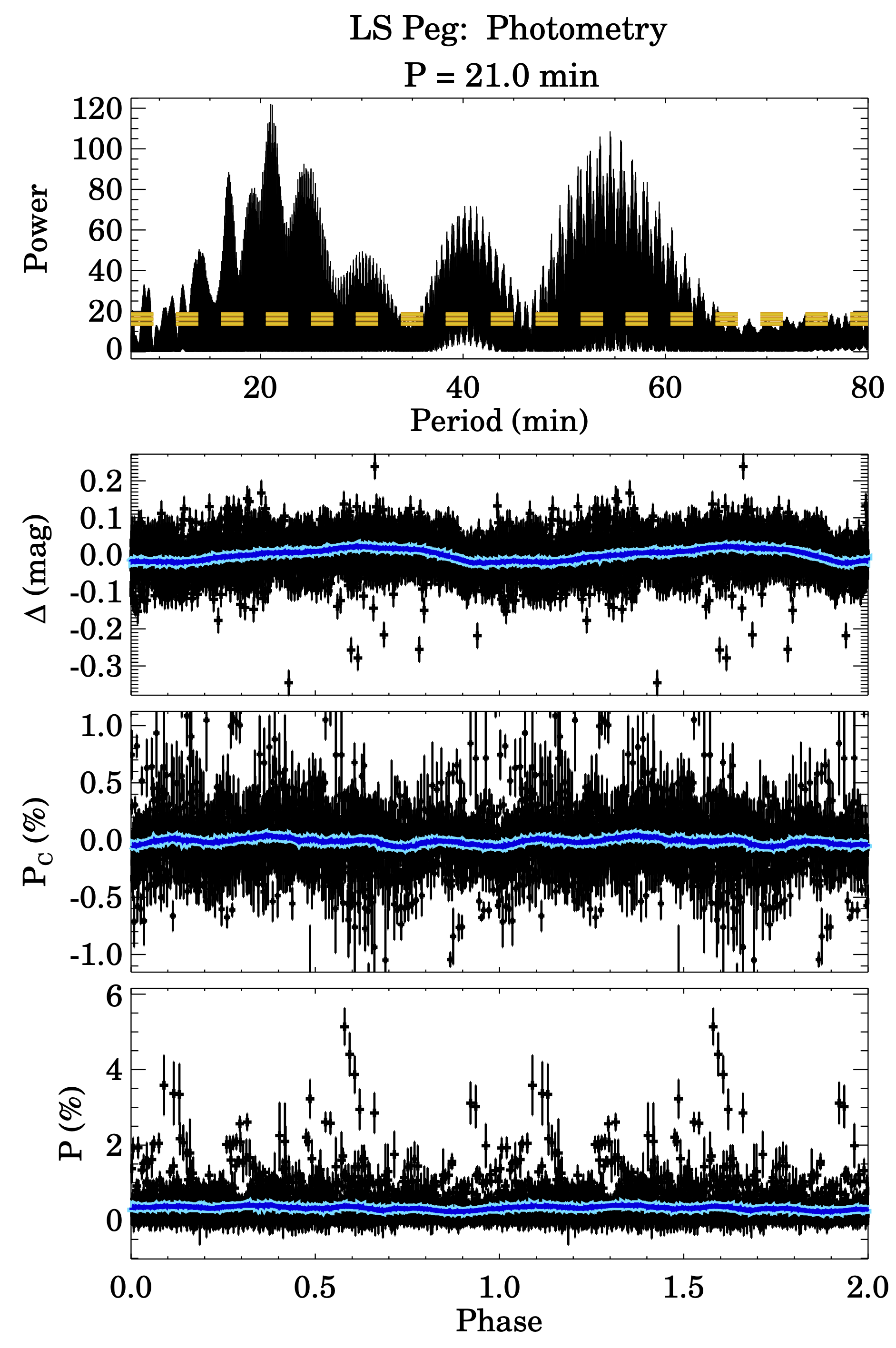}{0.45\textwidth}{}
          \fig{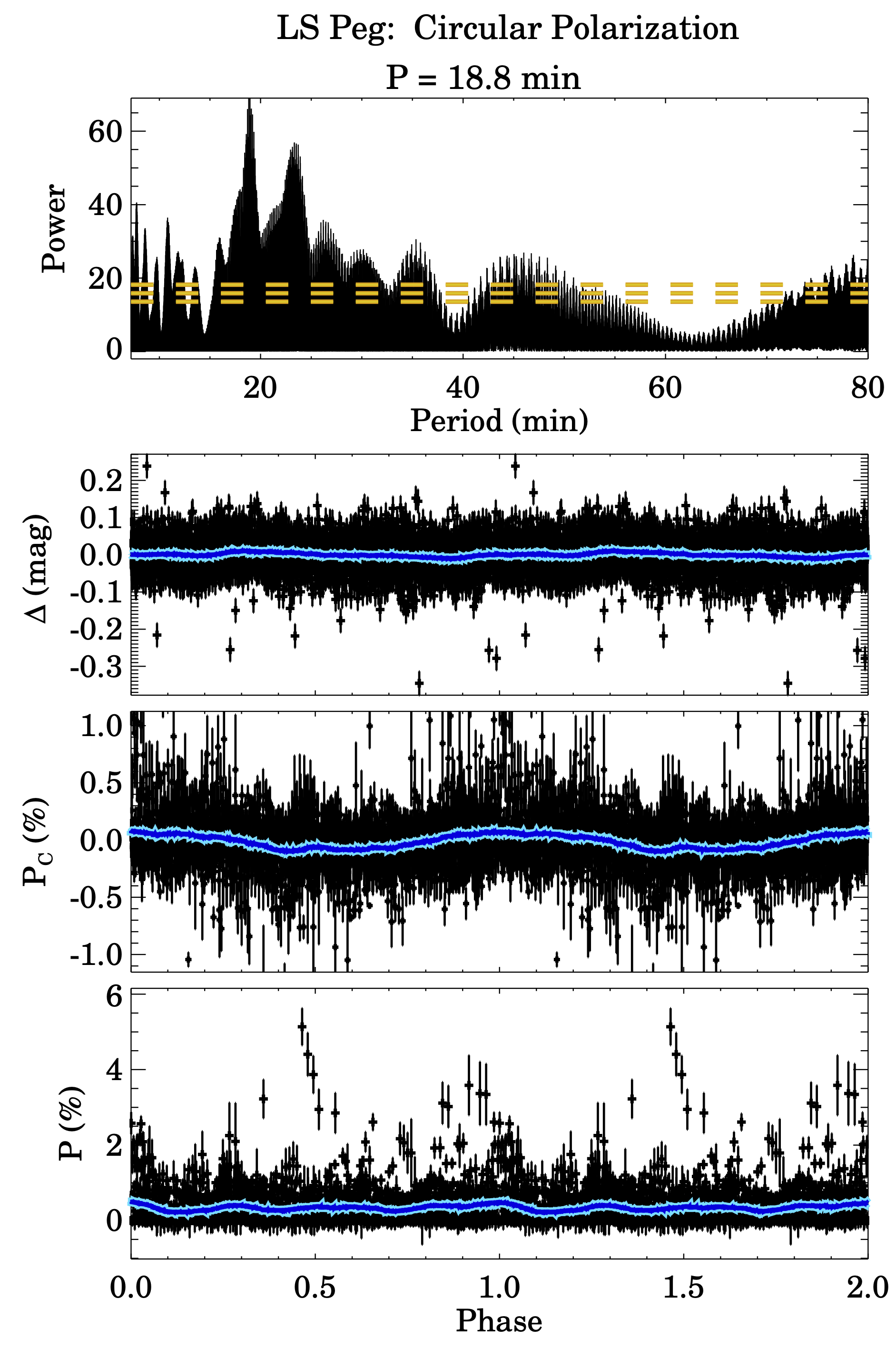}{0.45\textwidth}{}
          }
\caption{\added{Time-series analysis of LS~Peg.} The first column shows, from top to bottom, the power spectrum of the photometric data in \added{the} V and R$_{\rm C}$ bands and the phase diagrams of photometry, circular and linear polarimetric data, folded on the period P~=~21~min and arbitrary zero phase. The second\ column has the same structure of the first one, but the power spectrum is calculated for the circular polarization and has a peak at 18.8~min. The lines are colour-coded as in Figure~\ref{FigBO_Cet.1}.
\label{FigLSpeg.1}}
\end{figure*}

\deleted{The power spectrum of the light curve combining all bands is highly structured, hence the period error determination as the FWHM of a Gaussian fit does not perform so well. Therefore, in this case, the period error was estimated by the width of the envelope defined by the adjacent peaks in the periodogram.}
\replaced{The photometric data set of each night was filtered to remove low-frequency modulations and subtracted from the mean magnitude.}{The mean magnitude was subtracted from the photometric data set of each night, which were also filtered to remove low-frequency modulations.} The resulting power spectrum exhibits the strongest peak at 21.0~$\pm$~\replaced{1.7}{1.2}~min (see top of the first column of Fig.~\ref{FigLSpeg.1}). 
This period is consistent night-by-night, except for two of them (see Table~\ref{table:lspeg}). However, these nights do have a secondary peak around 20~min. The light curve folded on the 21~min periodicity shows a sinusoidal modulation with a semi-amplitude of 0.02~mag.
Circular and linear polarization combining the two bands \added{and} folded on the same period do not show a coherent variability (see first column of Fig.~\ref{FigLSpeg.1}). 
The other ``high-frequency" peaks seen in the power spectra are 16.8 and 24.2\,min. A peak at 19.3\,min superimposed on the main peak is also present. \citet{Taylor_1999} \deleted{have} found a photometric period of 20.7~min in a photometric time series collected on 12 nights spread over 18 days. The modulation was not stable in phase. Modulations around 19\,min and 16\,min have already been reported by \replaced{\cite{Garnavich_1992, Szkody_2001}}{\cite{Garnavich_1992} and \citet{Szkody_2001}}.

\begin{deluxetable}{ccc}
\tablecaption{Periods found in \added{the} LS~Peg data set.\label{table:lspeg}}
\tablewidth{0pt}
\tablehead{
\colhead{Data \added{type}} & \colhead{Date Obs.} &  \colhead{Period}\\
\colhead{} & \colhead{} & \colhead{(min)}  
}
\startdata
& All nights & 21.0~$\pm$~\replaced{1.7}{1.2} \\
& 2010 Oct 06 & 21.4~$\pm$~4.0\\
{Photometry} & 2010 Oct 12 & 19.7~$\pm$~1.7\\
& 2016 Oct 19 & 40.3~$\pm$~6.7\\
& 2019 \replaced{Sept}{Sep} 10 & 23.3~$\pm$~3.7\\
& 2019 \replaced{Sept}{Sep} 12 & 13.9~$\pm$~1.4\\
\hline
& All nights & 18.8~$\pm$~\replaced{1.0}{0.005}\\
& 2010 Oct 06 & 19.0~$\pm$~1.6\\
Circular polarimetry & 2010 Oct 12 & 18.6~$\pm$~1.2\\
& 2016 Oct 19 & 44.4~$\pm$~5.8\\
& 2019 \replaced{Sept}{Sep} 10 & 18.4~$\pm$~3.4\\
& 2019 \replaced{Sept}{Sep} 12 & 17.8~$\pm$~2.1\\
\enddata
\end{deluxetable}

Circular polarization data were not prewhitened. The periodogram of the entire data set shows a sharp peak centred at 18.8~$\pm$~\replaced{1.0}{0.005}~min (see top of the second column of Fig.~\ref{FigLSpeg.1}). This same period is also seen if we combine all 2010 data or all 2019 data. 
This suggests that LS~Peg has a stable periodicity in circular polarization along a baseline of 9~years. As in photometry, the main peak in the periodogram of 2016~\replaced{Oct.}{October}~19 occurs at around 40~min (see Table~\ref{table:lspeg}).
The circular polarization curves folded on 18.8~min exhibit positive and negative excursions with semi-amplitudes of 0.05\% (blue line in third panel of second column of Fig.~\ref{FigLSpeg.1}). 
The photometry and linear polarization do not show clear modulations at this period (see last panel). 

The periodograms of the linear polarization data show strong signals at low frequencies. Hence, we prewhitened the data and found a peak at 11.5~$\pm$~0.1~min, which could be related to half of the photometric 21~min period. A possible origin of this modulation  could be the reflection (scattering) of the \replaced{post-shock}{PSR} region emission in the inner regions of the disk.

As stated above, \cite{rodriguez/2001} also detected modulated circular polarization. They found a period of 29.6~min using 19 circular polarimetry spectra with a time resolution of about 10~minutes, which results in a Nyquist frequency of around 20~min. Therefore, their data were not adequate to search for periodicities of the order of those previously found in photometry. 

The periodic signals found in LS~Peg (Table~\ref{table:periods}) can be interpreted in terms of WD spin, QPOs, orbital period, or the combination of those as beat periods. Since the best explanation for the presence of circular polarization is cyclotron emission from a \replaced{post-shock}{PSR} region on the WD surface, the circular polarization modulation must be attributed to the rotation of a magnetic WD. Hence, our results indicate that the period of $\sim$19~min is the spin period of the WD, consistent with previous suggestions  \citep[e.g.][]{Szkody_2018}. 
In such a scenario, the photometric 21~min can be, within the uncertainties, the beat period between the WD spin period and the orbital period. 
This interpretation is in line with the linear polarization modulation at around 11~min being caused by internal scattering, since beat periodicities are usually associated with reflection of internal sources on the accretion structures.

V795~Her has spectroscopic properties very similar to LS~Peg \citep{Taylor_1999,Martinez_1999}, indicating a similar mass accretion configuration. This object has non-null circular polarization modulated at a period of 19.54~min \citep{rodriguez/2002b}, close to the periodicity attributed to QPOs in LS~Peg. So, in a comparative way, this reinforces the presence of a magnetic WD in LS~Peg.

\subsection{UU~Aqr} \label{results_uuaqr}

The variability of UU Aqr was discovered by \citet{Beljawsky_1926}. More than \replaced{a half century latter}{half a century later}, \cite{Berger_1984} found that the object has a strong UV excess. 
Photometric monitoring performed by \citet{Volkov_1986} and  \citet{Volkov_2003} confirmed the UV excess and revealed that UU~Aqr has deep eclipses up to 2~mag and strong flickering outside the eclipse, establishing UU~Aqr as a CV. Using multiwavelength eclipse mapping, \cite{Baptista_1996} suggested that UU~Aqr is an SW~Sex object.
This was confirmed by \cite{Hoard_1998} based on spectroscopic features, even though the \replaced{He~II}{\heii} emission line is weak in comparison with H$\beta$ and the line absorption is deepest at orbital phase $\sim$0.8. UU~Aqr has an average magnitude outside the eclipse of 13.5~mag with an eclipse depth \replaced{between}{around} 1.4~mag \citep[see Fig.~1 in ][]{Baptista_2008}. In the high state, a bright spot on the outer edge of the disk changes the shape of the eclipse profile. 
Its orbital period is 0.163580487(2)\,d (Borges B. \& Baptista R., private communication).
Using data collected in 2000, \cite{Patterson_2005} found a period of 4.2~h that is \replaced{associated with}{attributed to the} superhump phenomenon. However, \cite{Bruch_2019} did not detect superhumps in data obtained in 2018 September, only regular variations with a period of about 4~days.

UU~Aqr was observed during five nights distributed between August and October 2009 (see Table \ref{tab_data}). The V filter was used in all observations and the exposure times range between 4~s and 15~s. The magnitude was stable during those nights with an average value of V~$\approx$~13~mag. The magnitude dispersion in our data is around 0.1~mag for most nights. We found a large correlation between the photometry and polarimetry of UU~Aqr and those of the field stars on some nights, so they were not considered in the analysis (see Table~\ref{tab_data}). We also removed the observations obtained during the eclipses.

\begin{figure*}
\gridline{\fig{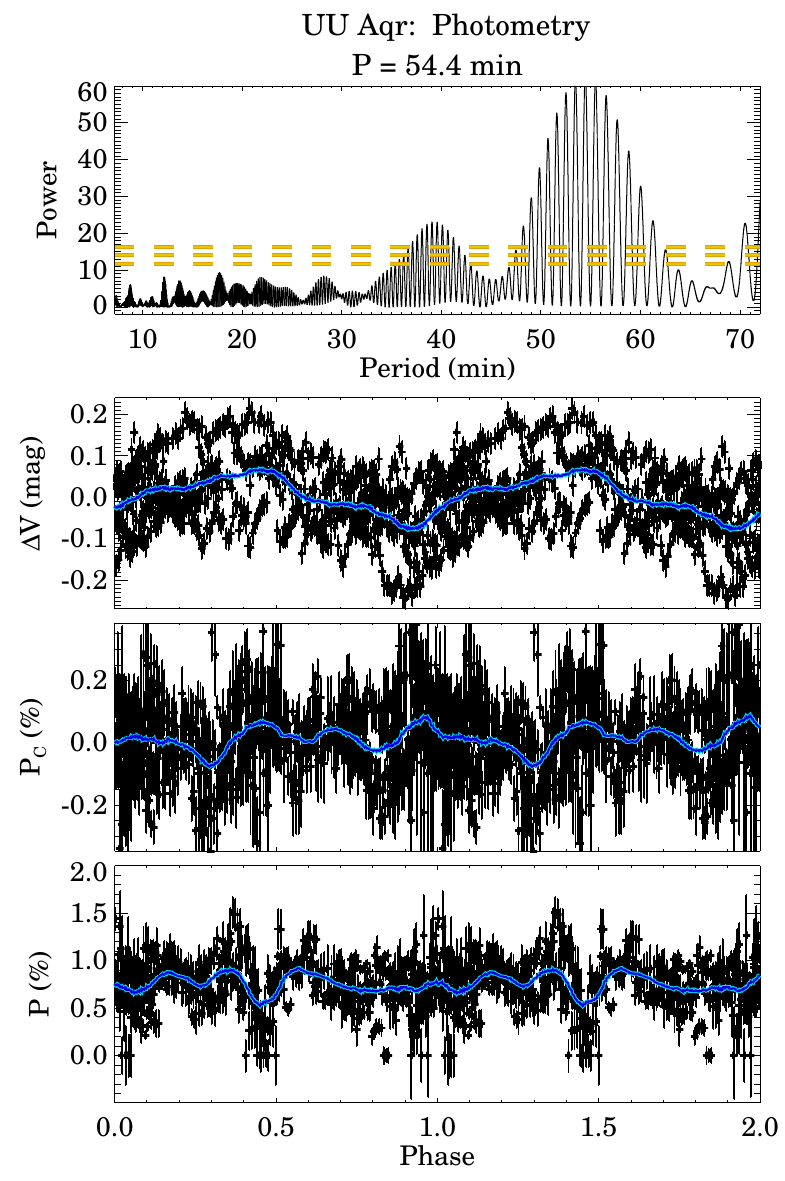}{0.4\textwidth}{}
          \fig{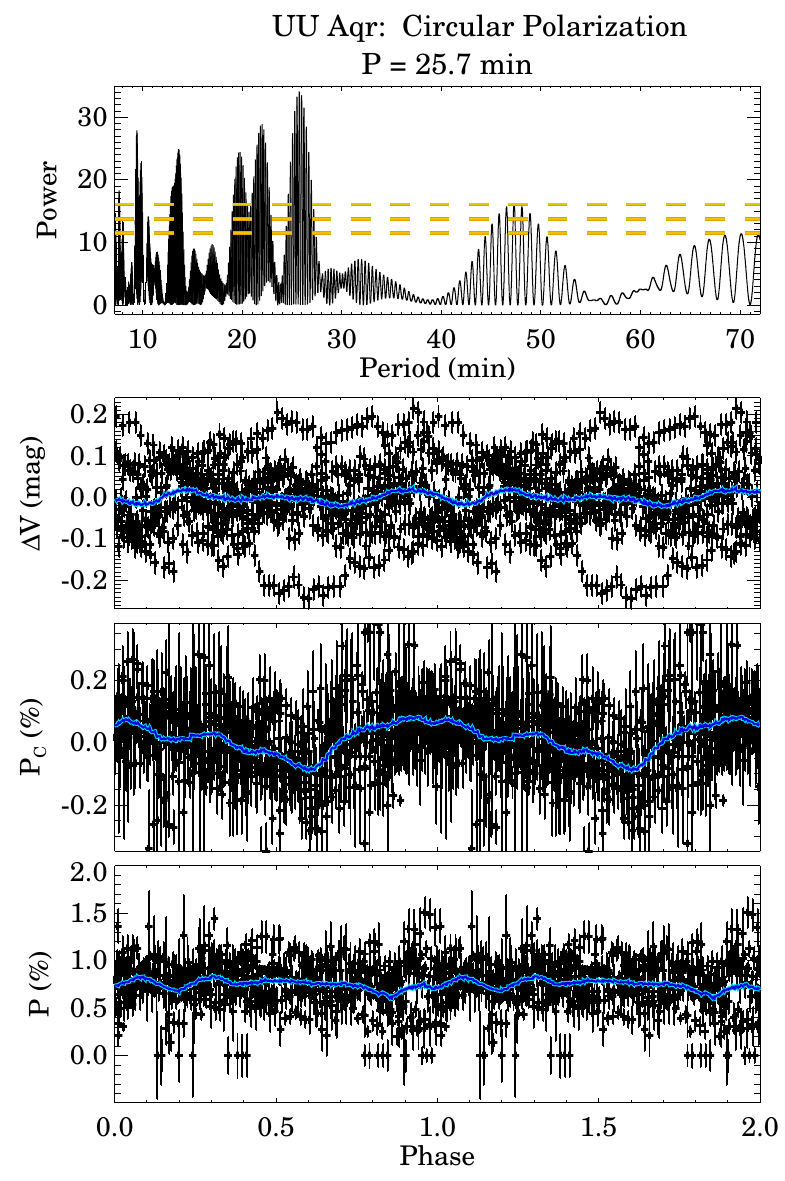}{0.4\textwidth}{}
          }
\caption{Time-series analysis of UU~Aqr. From top to bottom, the power spectrum of the photometry and the phase diagrams of photometry, circular and linear \replaced{polarisation}{polarization}. The first column shows the photometric results and the second column presents the period search in circular \replaced{polarisation}{polarization} data. The coloured lines are as defined in Figure~\ref{FigBO_Cet.1}.
\label{FigUUaqr.1}}
\end{figure*}

The power spectrum of \added{the} photometric data applying no filtering shows a strong peak at 54.4~$\pm$~\replaced{4.6}{0.5}\,min (Fig.~\ref{FigUUaqr.1}). The peak is also present if we subtract the orbital modulation from the data. The folded light curve displays a clear modulation with a semi-amplitude of around 0.2~mag. \deleted{The average folded polarimetric data suggests a regular pattern (see the blue line in the third panel of the first column of Fig.~\ref{FigUUaqr.1}).} The large dispersion around the average is due to the orbital flux modulation.
\added{The average folded polarimetric data suggests a regular pattern (see the blue line in the third panel of the first column of Fig.~\ref{FigUUaqr.1}).}

We did not subtract\deleted{ed} the low frequencies from the circular polarimetric data. The periodogram shows the strongest peak at 25.7~$\pm$~\replaced{0.5}{0.23}~min (see second column of Fig.~\ref{FigUUaqr.1}). This peak is also present in the periodograms of the individual nights, supporting the  persistent nature of this periodicity. The phase diagram of the circular polarimetry shows a modulation with peak-to-peak amplitude of 0.1\% (blue line). 
We did not find any statistically significant periods in the linear polarimetry.

As already discussed for the other objects, the only possible explanation for a periodic variability in circular polarization is the presence of a \replaced{post-shock}{PSR} region near the WD surface. Hence, the \deleted{$\sim$}26~min \added{periodicity} could be associated with the spin period of a magnetic WD. An argument in favour of this is the presence of a modulation with nearly twice this period in the photometry (Table~\ref{table:periods}). 
Another indirect evidence of a magnetic WD in UU~Aqr comes from the observation of variability on timescales of 0.5 -- 5~days during stunted outbursts \citep{Robertson_2018} who suggested it to be caused by ``blobby'' accretion associated with the fragmentation of the stream by the \added{WD} magnetic field.

In order to confirm the stability of the 26~min period and consequently the magnetic nature of the WD in UU~Aqr, it is necessary to obtain additional polarimetric data.
It is important to mention that the strong flickering of UU Aqr makes \added{it} difficult
to search for real periodic variabilities\replaced{ and can even produce fake peaks in the periodograms}{. The flickering is not polarized but, being a variable emission that dilutes the polarization, it can introduce noise in the polarization signal. It can also produce fake peaks in the total flux periodogram}.

\section{Discussion} \label{sec:discussion}

\subsection{Present status of circular polarization measurements of SW~Sex systems}
\label{disc:sw_pol}

\explain{We corrected the spelling of ''polarization'' in the section title above.}

We have performed polarimetry of 6 definite SW~Sex systems. In four of them (BO~Cet, SW~Sex, LS~Peg, and UU~Aqr), we have found \deleted{strong} evidence of modulated circular polarization. For V442~Oph and V380~Oph, the detection of circular polarization is uncertain.

Table~\ref{tab_list_hoard} summarises all circular polarization measurements of SW~Sex systems. This table is strongly based on information presented in The Big List of SW Sex Systems of \citet{hoard/2003}. We included the results of this paper and made some other changes, as follows. RR~Pic is quoted as polarized in The Big List, but this measurement refers to linear polarization, which could originate from other mechanisms than emission from the \replaced{PSR}{PSR region}. Therefore, it is not  evidence of magnetic accretion and, hence, it is not included in \deleted{the} Table \ref{tab_list_hoard}. We have also changed the ``polarized" classification of AO~Psc from ``N?" to ``Y?", since it has some evidence of modulation in circular polarization, but the errors are not small enough to confirm it. This object is also classified as an IP \citep[e.g.][]{Butters_2009}\footnote{There are other\deleted{s} objects in The Big List that are also classified as IPs. Some of them will be cited later in this discussion.}.
As a result, our table lists 27 systems, including BO~Cet and UU~Aqr for which the first measurements are presented here. 
The first lines of the table group the objects having confirmed or possible non-null circular polarization. 

A large fraction of the polarimetric observations of SW~Sex systems was performed by \cite{Stockman_1992}. Their observations consist of \replaced{only}{a single or} a few measurements\replaced{with integration times of around 8\,min for each polarimetric measurement. This}{. Each measurement is the result of data taken over an interval of 8\,min (or even longer), which} is inadequate to detect the circular polarization in these objects because the timescale of the modulation is usually of \replaced{the}{this} same order, causing the smearing of an intrinsically low polarization signal\added{, when present}. Hence, their negative detections are unreliable. Some other measurements in \replaced{this table}{Table~\ref{tab_list_hoard}} classified as ``polarized = N?" refer to long integrations with the same caveat of \cite{Stockman_1992}'s measurements. High signal-to-noise \added{ratio (polarization error smaller than 0.1\%)} and time-resolution of around 1\,min are necessary to \replaced{confirm the presence of}{detect} circular polarization in SW~Sex systems.

According to \replaced{Hoard's list}{The Big List}, there are 73 objects classified as SW Sex. They are divided into 30 definite, 18 probable, and 25 possible candidates. Table \ref{tab_list_hoard} lists 11 objects with some evidence of non-null circular polarization: this results in 15\% of SW~Sex systems with evidence of magnetic accretion from polarimetry. This is a non-negligible \replaced{group number}{fraction} since few objects have been observed with enough sensitivity and time resolution.
If we consider only the definite SW~Sex members, we obtain 33\% (10/30) of possible polarized objects. Taking into account only objects with confirmed modulated circular polarization, this fraction decreases to 23\% (7/30).

The above numbers \replaced{ascribe}{may reveal the presence of} magnetic WDs \replaced{to}{in} a considerable fraction of SW~Sex stars. More \replaced{high-SNR and time-resolved polarimetric observations}{polarimetric observations like the ones presented in this paper} are in demand to confirm previous measurements and expand the sample of observed SW~Sex objects.

\subsection{Periodicities in asynchronous magnetic CVs} \label{Periodicities}

\explain{In the section title, assynchronous was replaced with asynchronous.}

In this section, we discuss the main findings related to rapid variability in SW~Sex and compare them with what is observed in IPs. 

SW Sex systems usually show complex photometric variability. In some systems, periodicities related to the WD rotation, orbital cycle, or sidebands
have been found, similar to what is observed in some IP systems
\citep[e.g.][]{warner1986,Norton_1996}. Superhumps, with periods slightly longer or shorter than \replaced{P$_{\rm orb}$}{the orbital period}, are also present in some systems and are attributed to the precession of an eccentric accretion disk or disk \replaced{warp}{warping}. Another photometric variability common in the SW Sex class is quasi-periodic oscillations with time scales of 1000\,s. \citet{Patterson_2002} present a review of these different kinds of variability and propose that some of them can be explained by a magnetic WD. 

Non-orbital emission line periodicities are found in IPs as well as in SW~Sex stars. They can be observed as \added{a} modulation of the total emission line flux or as ``flares" in the line wings. In IPs, they are associated with the WD spin or the beat of \added{the} WD spin and orbital period. Examples of detailed studies of spin-phase resolved spectroscopy of IPs are \citet[FO~Aqr]{Marsh_1996} and \citet[DQ~Her]{Bloemen_2010}. Doppler tomography phased \replaced{to}{on} the spin cycle of seven IPs is presented by \citet{Hellier1999}. 
V1025~Cen is an\deleted{other} example of the complex variability behaviour in IPs \citep{Buckley_1998}, \replaced{showing that}{for which} the interpretation of the observed periods and the relation between values found in photometry and spectroscopy \replaced{is} {are} not straightforward.
Rapid spectroscopic variability has been found in 8 definite and 2 possible SW~Sex systems: 
EX~Hya \citep{Kaitchuck_1987}, which is classified as a possible SW~Sex system by \citet{hoard/2003};
LS~Cam \citep{Dobrzycka_1998};
BT~Mon \citep{Smith_1998};
LS~Peg \citep{rodriguez/2001};
V533~Her \citep{rodriguez/2002};
BO~Cet and V380~Oph \citep{rodriguez/2007a};
V1084~Her \citep{rodriguez/2009};
DW~UMa \citep{dhillon/2013}; 
and
SDSS~J075653.11+085831.8, in which the flaring is visible but no periodicity was found \citep{Tovmassian_2014}.
The periods range from 16 to 50\,min. Some authors suggested that these periods could be associated with the WD spin rotation, \replaced{analogous to what is observed}{as}  in IPs.

The literature has several examples of modulated circular polarization in IPs (e.g. \citealt{Piirola_2008} - V405~Aur - and \citealt{Katajainen_2010ApJ} - UU~Col). The periods are \replaced{P$_{\rm spin}$}{the spin period} or half this value. The last case is in fact explained by two maxima per 
\added{spin} cycle, hence the modulation of the polarization in IPs is always on the WD rotation. PQ~Gem, for instance, presents both behaviours \added{(one or two polarization peaks per cycle)}, depending on the band \citep{Potter_1997}.
Moreover, the total flux and the polarization do not necessarily vary in the same way.
NY~Lup and IGR~J15094$-$6649 have their polarization modulated with the spin period. However, the photometry of NY~Lup shows sideband periods in some bands, while the flux of IGR J15094$-$6649 modulates \deleted{only} with the WD spin \citep{Potter_2012}.

The SW~Sex systems with periodic modulation in \added{emission-line} flaring and circular polarization are V1084~Her, LS~Peg, BO~Cet, and V380~Oph (the last two \replaced{using}{considering} circular \replaced{measurements}{polarization} from this work).
\citet{rodriguez/2009} detected a period of 19.4~$\pm$~0.4~min in the \added{circular} polarization of V1084~Her (RX\,J1643.7+3402)
and twice this period in the radial velocity and equivalent width of \added{the} Balmer and \replaced{He~II~4686~\AA\ }{\heii} emission lines. The authors discussed that these observations can be understood in two ways\deleted{, but they favoured the first explanation}: (1) $\sim$19~min is half of the beat between \added{the} spin and orbital periods and $\sim$39~min is the beat period itself or (2) $\sim$19~min is half the spin period and $\sim$39~min is the spin period.
\added{They favoured the first explanation.}
\citet{Patterson_2002} obtained \added{a} long photometric time series on more than 50 nights. The power spectrum shows a QPO broad bump with a superimposed narrow peak at 17.38~min, which is consistent with the beat of 19~min and the orbital period. X-ray observations \deleted{clearly} show \replaced{an}{a clear} orbital modulation and a possible periodicity of around 26\,min \citep{Worpel_2020}, at odds with any previously reported period in this range. 

LS~Peg is one of the objects included in this work. The periods already claimed in the literature for this object as well as those found in this work are presented in Table~\ref{table:periods}. We could not confirm the previous claim of circular polarization modulated at 29.6~$\pm$~1.8~min \citep{rodriguez/2001}. Instead, we detected a modulation with a period of 18.8~$\pm$~\replaced{1.0}{0.005}~min\replaced{, that is in the same range in which photometric variability is systematically found}{, which is consistent with the periods systematically found in photometry} (see Table~\ref{table:periods}). We also found photometric periods in the interval \replaced{of}{between} 16.8 and 24.2~min. The main photometric period in our data is 21.0~$\pm$~\replaced{1.7}{1.2}\,min, consistent with the 20.7~$\pm$~0.3\,min from \citet{Taylor_1999}, and also consistent with the beat between our polarimetric period and the orbital period. For V1084~Her, the photometric period is also consistent with the beat of the polarization and orbital \replaced{cycle}{periods}, but it is larger than the polarimetric period, inversely to what is observed in LS~Peg. The spectroscopic period of 33.5~$\pm$~2.2\,min \citep{rodriguez/2001} is around \deleted{(but not exactly)} twice the polarimetric period, analogous to V1084~Cen. 

The results for V1084~Her and LS~Peg suggest a possible relation between the periods of \added{the emission-line} flaring and circular polarization, with the latter half the value of the \replaced{flaring}{former}. 
BO~Cet follows the same trend: a spectroscopic period of approximately 20\,min \citep{rodriguez/2007b} and a circular polarization period of 11.1~$\pm$~\replaced{0.3}{0.08}\,min. 
For V380~Oph, a similar situation happens.
We raised two questionable periods of 22 and 12\,min. Considering the first one, the same relation between circular polarization and spectroscopic flaring 
\citep[46.7~$\pm$~0.1\,min,][]{rodriguez/2007a} would be present.  
In both cases, the period found for spectroscopic flaring is also present in our photometric data, i.e\added{.} they have photometric periods \added{of} twice the polarimetric period. The same approximate relation is seen for UU~Aqr, for which no flaring was reported in the literature.

On the other hand, SW~Sex and V442~Oph have polarimetric periods (uncertain for V442~Oph) \added{of} approximately twice the photometric periods. Rapid spectroscopic variability was not reported for these two objects.

Magnetic WDs in spin-rate equilibrium can have different types of accretion flow \replaced{,  according to \cite{Norton_2008}:}{depending on the ratio between the WD rotation period, P$_{\rm spin}$, and the orbital period, P$_{\rm orb}$ \citep{Norton_2008}. 
Considering a mass ratio of 0.5,
the conditions for the different accretion geometries are the following:}
\added{if} P$_{\rm spin}$/P$_{\rm orb}$~$\lesssim$~0.1\added{, the geometry} will be \replaced{disklike}{disk-like}; 
\added{if} 0.1~$\lesssim$~P$_{\rm spin}$/P$_{\rm orb}$~$\lesssim$~0.6\added{,   it} will be \replaced{streamlike}{stream-like}, 
and \added{if} P$_{\rm spin}$/P$_{\rm orb}$~$\sim$~0.6\added{, it} will be \replaced{ringlike}{ring-like}\deleted{, considering the mass ratio of 0.5}. 
In \replaced{each case}{all cases}, the material is propelled in order to maintain angular momentum balance. The magnetic scenario for SW~Sex stars proposed by \cite{rodriguez/2001} has a magnetosphere radius extending up to the corotation radius with a corresponding relation between P$_{\rm spin}$ and P$_{\rm orb}$ given by\replaced{Equation~\ref{eq:rodrigues}. In this model, the gas stream from the secondary star overflows the disk and hits the magnetosphere of the primary}{:} 

\begin{equation}
\label{eq:rodrigues}
P_{\rm spin} \approx 0.31f^{3/2}P_{\rm orb}, 
\end{equation}
where $f$ is the corotation radius in units of R$_{L_{1}}$, which is the distance between the inner Lagrangian point L$_{1}$ and the \replaced{white dwarf}{WD}. 
\added{In this model, the gas stream from the secondary star overflows the disk and hits the magnetosphere of the primary.}

Table~\ref{table:Rodriguez-Gil} shows the results of the relation P$_{\rm spin}$/P$_{\rm orb}$ and $f$ for our sample of objects, considering the periodicities found in this paper and interpreted as the P$_{\rm spin}$. BO~Cet, V442~Oph, V380~Oph, LS~Peg and UU~Aqr can be classified as \replaced{disklike}{disk-like} and their $f$ values are consistent with the interval of 0.4~--~0.6~R$_{L_{1}}$ \citep{Groot_2001}. Only SW~Sex itself exhibits the accretion flow as \replaced{streamlike}{stream-like}, and a corotation radius of 0.7~R$_{L_{1}}$,  slightly larger than the value measured by \cite{Groot_2001}. 

\begin{deluxetable}{ccccc}
\tablecaption{Relations of orbital period and spin period considering our results.\label{table:Rodriguez-Gil}}
\tablewidth{0pt}
\tablehead{
\colhead{Object} & \colhead{P$_{\rm orb}$} &  \colhead{ P$_{\rm spin}$} & \colhead{P$_{\rm spin}$/P$_{\rm orb}$} & \colhead{$f$}
\\
\colhead{} & \colhead{(min)} & \colhead{(min)}  &  & \colhead{(R$_{L_{1}}$)}
}
\startdata
BO Cet & 201.36 & 11.1 & 0.06 & 0.32 \\
SW Sex & 194.31 & 41.2 & 0.21 & 0.78 \\
V442 Oph & 179.07 & 19.4 & 0.11 & 0.50 \\
V380 Oph & 221.91 & 12.7 & 0.06 & 0.32  \\
LS Peg & 251.67 & 18.8 & 0.07 & 0.39 \\
UU Aqr & 235.56 & 25.7 & 0.11 & 0.50 \\
\enddata
\end{deluxetable}

\subsection{Inclination versus polarization} \label{Inclination}

The SW~Sex class was initially supposed to be composed only of eclipsing systems. In spite of the discovery of an increasing number of non-eclipsing systems, \replaced{the fraction of eclipsing SW~Sex (more than a half) remains larger than that expected for a homogeneous distribution of inclinations.}{more than a half of the SW~Sex systems are eclipsing, which is not consistent with a homogeneous distribution of inclinations.} 
\added{This high incidence of eclipsing systems allows us to verify if the detection of polarization in SW~Sex objects is correlated with their inclination.}
For the present discussion, it is not relevant if this is an observational or historical bias or a physical characteristic of the SW~Sex phenomenon. \deleted{Below, we verify if the detection of polarization in these systems is correlated with their inclination.}

Notwithstanding the small number of objects, Table~\ref{tab_list_hoard} shows that \deleted{the} 73\% (8/11) of SW~Sex stars with possible circular polarization are non-eclipsing systems. If we focus only on objects with confirmed circular polarization, 4 out of 7 are non-eclipsing. Therefore, there is a higher incidence of polarized objects among non-eclipsing systems. 

A possible origin for such a correlation could be the following. In the magnetic accretion scenario, the cyclotron emission is produced very near the WD surface \added{and is responsible for the observed polarization}. 
The \replaced{geometrically thick}{accretion} disks of SW~Sex systems could block the direct view of the \replaced{WD surface}{PSR region} in high-inclination systems, which would prevent us from observing \replaced{a}{the} polarized component in the total system emission at all phases.
\added{Such occultation is favoured by geometrically thick or flared disks, which could explain some spectroscopic features of SW Sex systems \citep[see][and references therein]{dhillon/2013}. In fact, models of the vertical structure of accretion disks predict an increase of the disk thickness with $\dot{M}$ \citep{meyer1982}.}
On the other hand, the \replaced{emission}{flux} from the disk increases with its  projected area\replaced{ making}{, which makes} the dilution of a possible \replaced{post-shock}{PSR} region emission larger for \replaced{smaller}{lower} inclinations. Hence, there should be an optimal inclination where the detection of the polarization would be most favoured. This inclination is likely near the maximum inclination for which no eclipse is seen.

\subsection{SW~Sex stars in X-rays} \label{X-rays}

\replaced{The post-shock region of IPs emits in X-rays and makes them strong sources at high energies.}
{In addition to the optical cyclotron emission, the material in the postshock region also cools by bremsstrahlung emission in X-rays. Hence, models of the optical and X-ray emission of magnetic CVs can be used to constrain their physical and geometrical properties \citep[e.g.][]{silva2013,oliveira2019}.}
The \added{X-ray} flux is usually modulated with the WD rotation due to variable absorption with the viewing angle or \replaced{self-occultation (when the region is occulted by the WD)  of the emitting region}{occultation of the emitting region by the WD}. 
\citet{Mukai_2017} presents a comprehensive review on X-ray emission from accreting WDs. In this section, we \replaced{give a quick}{briefly} overview \deleted{of} the X-ray emission of SW~Sex objects.

Table \ref{table:x-rays} shows the SW~Sex objects that have X-ray\deleted{s} counterparts.
They add up to 10 objects, corresponding to $\sim$14\% of the 73 systems of The Big List. \replaced{Only four of them have confirmed or suspected}{Five of them have reported} \added{X-ray} periodicity\replaced{. And from these objects, two}{ and two of them} 
are also classified as IPs: EX~Hya \citep{Hellier_2000} and AO~Psc \citep{Hellier_2005}. \added{The IP systems are not discussed here.}

Three \sss are X-ray\deleted{s} sources and have \replaced{confirmed}{positive} \added{circular} polarization \added{detection (see Table \ref{tab_list_hoard})}: LS~Peg, V533~Her, and V1084~Her. 
\added{\citet{baskill/2006} claimed a period of 30.9~min in the X-ray emission of LS~Peg, which is consistent with the circular polarization period of 29.6~min reported by \citet{rodriguez/2001}. However, }
\cite{Ramsay_2008} did not find any \replaced{period}{periodicity} in \replaced{LS~Peg}{further} X-ray \replaced{emission}{data} 
\added{\replaced{as}{and} we did not confirm the period of 29.6~min in circular polarization either}
(see also Section~\ref{sec_results_lspeg}). V533~Her and V1084~Her \replaced{were}{have been} recently studied by \citet{Worpel_2020} using XMM-Newton. V533~Her has low X-ray luminosity and its light curve exhibits an uncertain periodicity of 22.48~min, which is close to (but inconsistent with) \replaced{a}{the} 23.33~min \replaced{period obtained from}{periodicity detected in} the equivalent width of \deleted{the} emission lines by \cite{rodriguez/2002}.
The relation between these two periods cannot be explained by a beat with the orbital period. \added{There is only one measurement of the \deleted{optical} optical circular polarization of V533~Her, which corresponds to a marginal 3-sigma detection \citep{Stockman_1992}.}
\replaced{In V1084~Her, a periodicity of around 26~min was found, but it}
{\citet{Worpel_2020} reported an X-ray periodicity of around 26~min for V1084~Her, which}
does not correspond to the optical circular polarization period 
\added{of 19.4~min reported by \citet{rodriguez/2009}}
or \added{to} any other periodicity reported from optical photometry or spectroscopy.

From the above discussion, there is no \added{clear} evidence of magnetic accretion \replaced{in}{from} the X-ray light curves of \ssss. \added{In particular, the modulations observed in X-rays are not seen in optical measurements.} \citet{Patterson_2002} discusse\replaced{s}{d} if the absence/small levels of X-ray emission is inconsistent with the magnetic scenario. They proposed that the high density in the accretion column could prevent the \added{production of a} shock, \replaced{and hence the gas, heating to keV temperatures}{which hinders the gas to reach keV temperatures}. However, the subject is far from being settled. Hopefully, forthcoming X-ray\deleted{s} surveys (e.g. e-Rosita) will shed light on this topic.

\begin{deluxetable}{ccccc}
\tablecaption{SW~Sex stars detected in X-rays.
\added{If a periodic modulation is present in the data, the period, $P$, is shown in the second column.} \replaced{The}{A} question mark indicates uncertainty in the period.
\explain{The first line was added in this revised manuscript.}
\label{table:x-rays}}
\tablewidth{0pt}
\tablehead{
\colhead{Object} & \colhead{P} &  \colhead{Reference}
\\
\colhead{} & \colhead{(min)} & \colhead{} 
}
\startdata
LS Peg & 30.9 $\pm$ 0.3 & \cite{baskill/2006}\\
LS Peg & -- & \cite{Ramsay_2008} \\
V533~Her &  22.48 (?) & \cite{Worpel_2020}\\
V1084~Her & 25.82 (?) & \cite{Worpel_2020}\\
AO Psc & 13.4 (P$_{\rm spin}$) & \cite{Johnson_2006}\\
AH Men & -- & \cite{White_2000}\\
EX Hya & 67 (P$_{\rm spin}$) & \cite{Heise_1987}\\
DW UMa & -- & \cite{Hoard_2010}\\
WX Ari & -- & \cite{White_2000}\\
PX And & -- & \cite{White_2000}\\
UX UMa & -- & \cite{Pratt_2004} \\
\enddata
\end{deluxetable}

\subsection{CV demography and SW~Sex systems}
\label{CV:demography}

\cite{Schwope_2018} and \cite{Pala_2020} have provided volume-limited studies of CVs,\ considering distances from the {\em Gaia} second data release. The latter \added{authors} provide\deleted{s} 
42 systems within \added{a distance of} 150 pc. This sample is dominated by systems with  low \replaced{mass-accretion}{mass-transfer} rate: it contains only \replaced{2 nova-like CVs and}{three nova-like CVs including} one SW Sex object (EX~Hya, which is compiled in that work as an IP). They found that 36\% of the objects host a magnetic WD. The sample is dominated by objects below the period gap (83\%), but the fractions of magnetic \replaced{CVs}{systems} below and above the gap are around the same\replaced{ (37\% and 28\%), particularly}{, 37\% and 28\% respectively, } considering the small number of objects. If the fraction of magnetic CVs does not depend on the \replaced{accretion}{mass-transfer} rate or \deleted{on} the orbital period\deleted{ (having a constant value for all types of CV)}, we would expect that around 36\% of the \replaced{nova-likes}{nova-like CVs} would harbor a magnetic WD. 
Here, \replaced{nova-likes}{nova-like variables} stand for \replaced{high-accretion}{high mass-transfer} rate CVs that are not polars or IPs. The fraction of \sss among the \replaced{nova-likes}{nova-like systems} in the \cite{Ritter_2003} catalog (version 7.24) is 46\%, which is numerically consistent with the assumption that \sss are \replaced{those}{the missing} magnetic \replaced{nova-likes}{nova-like CVs}.

\subsection{SW~Sex stars in the context of the evolution of cataclysmic variables} \label{CV:evolution}

The CVs in the orbital period range between 3 and 4 hours that have sufficient observations show at least some SW~Sex characteristics. A possible interpretation is that all long-period CVs have to evolve through this SW~Sex regime before entering the period gap \cite[e.g.][]{rodriguez/2007b, Schmidtobreick_2012}. In fact, the number of dwarf novae around 4~h decreases in comparison with the number of SW~Sex stars.
In this scenario, the SW~Sex stars represent an important stage in CV\deleted{s} evolution.

CVs evolve towards shorter orbital periods driven by angular momentum loss through magnetic braking and gravitational radiation \citep*[e.g.][]{Paczynski_1967,Verbunt_Zwaan_1981}. 
In addition to these mechanisms, there is additional angular momentum losses due to mass transfer itself, usually called consequential angular momentum loss \citep[CAML, e.g.][]{King_Kolb_1995}.
Reasonable candidates for this sort of angular momentum loss include nova eruptions and circumbinary disks, among others.
In the last couple of years, evidence supporting the importance of CAML in CV evolution \replaced{have}{has} been growing  \citep[e.g.][]{Schreiber_2016,Nelemans_2016,Liu_Li_2016}, especially regarding stability for dynamical mass transfer and the required extra angular momentum loss below the orbital period gap \citep[][]{Knigge_2011,Pala_2017}.
In particular, the empirical formulation by \citet{Schreiber_2016} provides an explanation for the paucity of helium-core WDs among the CV population \citep{Zorotovic_2011,McAllister_2019}, as well as 
the space density and the fractions of short- and long-period systems consistent with observations \citep[][]{Belloni_2020a,Pala_2020}.
In addition, it can also explain the properties of detached CVs crossing the orbital period gap \citep{Zorotovic_2016}, the existence of single helium-core WDs \citep{Zorotovic_2017}, and the mass density of CVs in globular clusters \citep{Belloni_2019}.

CAML might solve several problems of the CV evolution model, but not all of them. 
There are some inconsistencies between predictions and observations that are most likely connected to our ignorance of magnetic braking.
One prescription widely used in CV investigations is that proposed by \citet*{RVJ}.

A well-known problem of this prescription, which is directly related to this work, is the contradiction between the expected mass-transfer rates among SW~Sex stars (given the brightness of such systems) and the predicted ones \cite[e.g.][]{rodriguez/2007b}.
Indeed, SW~Sex stars cluster at the upper edge of the orbital period gap, between \replaced{$\sim3-4$~h}{$\sim3$ and $\sim4$~h}. 
According to the prescription by \citet*{RVJ}, the mass-transfer rate drops as a CV ages, which means that just above the orbital period gap, this prescription provides the lowest mass-transfer rates.
On the other hand, the mass-transfer rates expected from the characteristics of SW~Sex stars are supposed to be the highest among all CVs.

Other evidence\deleted{s} that the magnetic braking prescription should be different \replaced{was}{were} recently provided by \citet{Belloni_2020a} and \citet{Pala_2020}, who showed that the observed and predicted fractions of period-bouncers are in serious disagreement, with those predicted being much larger than observed.
In addition, these authors showed that the evolutionary trend in the WD effective temperature above the period gap, which is directly connected with the mean mass-transfer rate, is opposite to the observed one.
Finally, \citet{FuentesMorales_2020} showed that the predicted fraction of old novae just above the orbital period gap is much smaller than the observed fraction.
These four pieces of evidence point towards a needed revision of the magnetic braking prescription.
SW~\replaced{Sew}{Sex} stars are therefore important objects in this context, which can provide significant constraints on the CV evolution model, especially considering that they are more abundant than period bouncers and long-period CVs having WDs with measured effective temperatures.

We note, however, that there is some progress in this regard.
\citet{Knigge_2011} provide a list of magnetic braking recipes, which are illustrated in their Fig.~2.
Among them, the formulation developed by \citet{Kawaler_1988} clearly predicts an increase in the angular momentum loss rate when the CV evolves towards shorter orbital periods, which would, in turn, provide the largest mass-transfer rates just at the upper edge of the orbital period gap.
This could potentially explain the expected mass-transfer rates of SW~Sex stars, which could  make SW~Sex stars a natural stage in the long-period CV pathway. 
Additionally, the observed distribution of WD effective temperatures versus orbital period could be reproduced.
Belloni et al. (in preparation) upgrade\deleted{d} the prescription by \citet{Kawaler_1988} and show these and other issues related to magnetic braking might be solved.

\subsection{Origin of magnetic cataclysmic variables} \label{CV:origin}

There are currently two main scenarios that could account for the formation of magnetic WDs in close binaries.
In the first one, during common-envelope evolution, a dynamo driven by shear due to differential rotation in the hot outer layers of the degenerate core \citep{Wickramasinghe_2014} would be responsible for the magnetic field generation.
This idea was originally designed for single WDs, but was generalized by \citet{Briggs_2018}, who extended the model to the case in which the binary survives the common-envelope evolution.
However, \citet{Belloni_2020b} detected several flaws in this model. 
One of them is the difficulty in explaining the complete absence of hot and young WDs among the population of post-common-envelope WD $+$ late-type main-sequence stars \citep[e.g.][and references therein]{Parsons_2021}. 
In addition, the predicted fractions of magnetic WDs in the different populations of close binaries are much higher than the observed ones.

An alternative scenario for the origin of single low-field magnetic WDs was \deleted{recently} proposed by \citet{Isern_2017}, who argued that a dynamo similar to those operating in planets could be responsible for weak magnetic field generation in WDs.
This is possible because, once the WD temperature becomes low enough, its interior starts to crystallize, which, in turn, allows the generation of a magnetic field through a dynamo.
This scenario, if applied to close binaries, could help to explain the low-temperature WDs in the population of detached magnetic post-common-envelope binaries.
However, the magnetic fields predicted by \citet{Isern_2017} are much weaker than those observed in magnetic WDs among the different population of close binaries.

If all or part of \sss have high-magnetic moment \citep[e.g.][]{Patterson_2002},
any theory aiming to explain the origin of high-field magnetic WDs would need to account for their properties.
That said, SW~Sex stars are important pieces to constrain evolutionary models of CVs, and understanding whether they are predominantly magnetic or not, should provide useful constraints  to any model aiming to explain the origin and evolution of magnetic WDs in close binaries.

\section{Conclusions}  \label{sec:conclusions}

We reported time-series analyses of photometric and polarimetric data of six SW~Sex stars\added{: BO~Cet, SW~Sex, V442~Oph, V380~Oph, LS~Peg, and UU~Aqr}. We associated the periodicities found in circular polarization with the spin period of \replaced{the}{a} magnetic WD, based on the assumption that \replaced{it}{the polarized flux} is related to cyclotron emission \replaced{of the}{from a} \replaced{post-shock}{PSR} region. 
We found the following \added{tentative} periods from \added{circular} polarimetry: 11.1~min in BO~Cet, 41.2~min in SW~Sex, and 25.7~min in UU~Aqr. \replaced{Also, below the FAP levels of 1\%, we found the periods of 22.0~min for V380~Oph and of 19.4~min for V442~Oph.}{We also found uncertain periodicities (below the FAP level of 1\%) of 22.0~min and 19.4~min for V380~Oph and V442~Oph, respectively.} We confirmed the detection of circular polarization in LS~Peg, previously reported by \cite{rodriguez/2001}. However, differently from these authors, we found a \deleted{circular polarimetry} period of 18.8~min, which we assumed as the probable period of the WD rotation. 
Considering these new \added{possible} detections, 15\% of all SW~Sex in 
Hoard's Big List of SW~Sextantis Stars
(which contains 73 objects) have direct evidence of magnetic accretion from circular polarimetric data.

There is a weak indication that the circular polarimetric period is half of the emission-line flaring period. There is also a tendency of detection of circular polarization in non-eclipsing \added{SW~Sex} systems: 73\% of objects with detected polarization are non-eclipsing, contrasting with the trend of SW~Sex systems to show eclipses.

The recent finding that 36\% of the CVs in a volume-limited sample are magnetic \citep{Pala_2020} agrees with an interpretation that SW~Sex systems are the magnetic portion of the nova-like systems. If this is really true, the formation of magnetic CVs should also explain the SW~Sex systems, and not only polars and IPs. In fact, any model for the origin and evolution of CVs should also reproduce the tendency of SW~Sex systems to cluster just above the orbital period gap.

\added{Additional time-resolved and high signal-to-noise ratio circular polarimetric measurements of SW~Sex systems are necessary to confirm the suggested periods and to search for magnetic accretion in those objects.}

\acknowledgments

\added{The authors thank the referee for corrections and suggestions to the manuscript.}
This study was supported by grants \#2018/05420-1, \#2015/24383-7, \#2013/26258-4, Funda\c c\~ao de Amparo \`a Pesquisa do Estado de S\~ao Paulo (FAPESP).
C.V.R. is also grateful to the support of Conselho Nacional de Desenvolvimento Cient\'ifico e Tecnol\'ogico - CNPq (grant \#303444/2018-5). C.E.F.L. acknowledges a PCI/CNPQ/MCTI post-doctoral support. P.S. acknowledges support from NSF AST-1514737. 
A.S.O. acknowledges S\~ao Paulo Research Foundation
(FAPESP) for financial support under grant \#2017/20309-7. D.B. was supported by the grant {\#2017/14289-3}, S\~ao Paulo Research Foundation (FAPESP). S.S. was supported by the Slovak Research and Development Agency under the contract No. APVV-15-0458, by the Slovak Academy of Sciences grant VEGA No. 2/0008/17
and was partially supported by the Program of Development of M. V. Lomonosov Moscow State University ``Leading Scientific Schools", project ``Physics of Stars, Relativistic Objects and Galaxies".
The authors thank \replaced{to}{the} MCTI/FINEP (CT-INFRA grant 0112052700) and the Embrace Space Weather Program for the computing facilities at INPE. This work was based on observations made at Observat\'orio do Picos dos Dias managed by the Laborat\'orio Nacional de Astrof\'\i sica/MCTI, Brazil. We acknowledge the use of  observations from the AAVSO International Database contributed by observers worldwide and used in this research.
The authors also acknowledge Deon\'\i sio Cieslinski, for his help \replaced{on}{in} the beginning of the project, and Matthias Schreiber, for his comments on aspects of CV evolution.

%

\facilities{LNA:1.6m, SOS/SAS, AAVSO}


\software{IRAF \citep{Tody/1986, Tody/1993}}



\clearpage

\appendix

\section{Polarimetric Measurements of SW Sex Systems} \label{app:sw_pol}

In this section, we present a table with \replaced{compiled information about polarisation}{ a compilation of polarization} measurements of SW~Sex systems \replaced{in}{from} the literature\replaced{ and}{. The table also includes the measurements presented } in this work\replaced{. It}{ and} is strongly based on the information presented in \replaced{\citet{hoard/2003}'s}{The} Big List of SW Sex Systems \added{\citep{hoard/2003}}.

\begin{longrotatetable}
\begin{deluxetable*}{ccccccccc}
\tablecaption{Summary of previous circular polarization measurements of SW Sex systems listed in \citet{hoard/2003}. The first objects are those having confirmed non-null polarization. \label{tab_list_hoard}}
\tablewidth{700pt}
\tabletypesize{\scriptsize}
\tablehead{
\colhead{Object} & \colhead{SW Sex status} & 
\colhead{Eclipse} & \colhead{Polarized} & 
\colhead{$P_c$} & \colhead{Wavelength range} & 
\colhead{Period} & \colhead{Reference} \\ 
\colhead{} & \colhead{} & \colhead{} & \colhead{} & 
\colhead{(\%)} & \colhead{} & \colhead{(min)} &
\colhead{} 
} 
\startdata
UU~Aqr      & Definite   & Deep &  Y  & -0.1 -- +0.1    &  V & 25.7~$\pm$~\replaced{0.5}{0.23} & This paper \\     
BO Cet     &  Definite    & None &    Y       & -0.1 -- +0.1  &  V, R$_{\rm C}$ &11.1~$\pm$~\replaced{0.3}{0.08} & This paper \\
RR Cha      & Definite & Deep & Y & 5 -- 10   & 3500--9000~\AA    & \nodata & \citet{rodriguez/2003} \\
V533 Her    & Definite & None & Y? & -0.38(11) and 0.00(7)  & 3200--8600~\AA  & \nodata & \citet{Stockman_1992} \\
V795 Her    & Definite & None & Y & 0.1 -- 0.22   & U   & 19.54 &  \citet{rodriguez/2002b} \\
V1084 Her   & Definite & None & Y & -0.5 -- +0.5 & 3600--5350~\AA    & 19.4~$\pm$~0.4 &  \citet{rodriguez/2009} \\
V380 Oph    & Definite & None & Y? & -0.1 -- +0.1  & V    & 22.0~$\pm$~\replaced{2.2}{1.2} & This paper \\
  &  &  & N? &  +0.12(16) & 3200--8600~\AA &  \nodata  & \cite{Stockman_1992} \\
V442 Oph    & Definite & None & Y? &   -0.1 -- +0.1      & V, R$_{\rm C}$ &  19.4~$\pm$~\replaced{1.7}{0.4}   & This paper \\
    &  & & N? &  +0.01(5) and -0.03(8)         & 3200--8600~\AA &  \nodata   & \cite{Stockman_1992} \\
LS Peg      & Definite & None & Y & 0.05 -- +0.05 & V, R$_{\rm C}$   & 18.8~$\pm$~\replaced{1.0}{0.005} & This paper \\
 &  &      &     Y       &  0.0 -- -0.3  & 3720--5070~\AA  &  29.6~$\pm$~1.8  &  \citet{rodriguez/2001} \\
AO Psc      & Possible & None  & Y? & 1.3~$\pm$~0.5 & I  & 13.42   & \citet{Butters_2009} \\
SW Sex      & Definite & Deep & Y & -0.2 -- +0.2 &  R$_{\rm C}$   & 41.2~$\pm$~\replaced{7.7}{8.5} & This paper \\
     &  &  &N? &  +0.05(14) and -0.03(5)   & 3200--8600~\AA &  \nodata  & \cite{Stockman_1992} \\ 
\\
PX And      & Definite & Shallow & N? & -0.02(9) and +0.06(4)          &
3200-8600~\AA  &    & \cite{Stockman_1992} \\
TT Ari      & Definite & None & N? & -0.01(2) and -0.02(2)          & 
3200--8600~\AA &     & \cite{Stockman_1992} \\
OZ Dra      & Definite & Deep & N? & $<$0.24(17)                 &
4000--8000~\AA &     & \citet{Wolfe_2003} \\
EV Lyn      & Possible & Shallow  & N? & $<$0.1--0.4                  &
--            &      & \citet{Szkody_2006} \\
EX Hya      & Possible & Shallow  & N? & -0.02(4) and +0.01(2)          &
5700--9200~\AA    &  & \citet{Stockman_1992} \\
BT Mon      & Definite & Deep  & N? &  0.10(17), 0.04(6) and 0.21(12)& 3200--8600~\AA  &   &  \citet{Stockman_1992} \\
V Per       & Probable & Deep   & N? & -0.81(82)                    &
3200--8600~\AA &     & \cite{Stockman_1992} \\
V347 Pup    & Probable  & Deep  & N? & $<$0.5                     & 0.34--1.6~$\mu$m & & \citet{Buckley_1990} \\
V348 Pup    & Definite & Deep & N? & -0.09(11) and +0.14(13)       &
5500--8400~\AA  &    & \citet{Tuohy_1990} \\
VZ Scl      & Possible & Deep  & N? & +0.18(8)                     &
3500--9200~\AA  &    & \citet{Cropper_1986} \\
LX Ser      & Probable  & Deep & N? & +0.04(7)                     &
3200--8600~\AA &     & \citet{Stockman_1992} \\
WY Sge      & Possible & Deep  & N? & +0.32(46)                    &
3200--8600~\AA  &    & \cite{Stockman_1992} \\
DW UMa      & Definite & Deep & N? &  +0.06(14)                   &
3200--8600~\AA &     & \cite{Stockman_1992} \\
UX UMa      & Probable & Deep   & N? & -0.06(4)                     &
3200--8600~\AA    &  & \cite{Stockman_1992} \\
1RXS J064434.5+334451 & Probable & Deep  & N       & +0.003(3)                    & 4200--8400~\AA &     & \citet{Sing_2007} \\
SDSS J210131.26+105251.5 & Possible & None?  & N? & $<$0.4                      &
V                &  & \citet{Homer_2006} \\
\enddata
\tablecomments{\added{Column 5 shows the measured percentage of circular polarization. Errors are shown in parentheses, when available.}}
\end{deluxetable*}
\end{longrotatetable}

\section{Improving the data reduction of the IAGPOL circular polarization mode}
\label{app:k}

This section explains a procedure \replaced{in}{for} the data reduction of the circular polarimetric mode of IAGPOL \citep{Magalhaes_1996}. The procedure takes into account the expected value of the ratio of the sum of the ordinary counts of all images used to calculate one polarization point relative to the sum of the extraordinary counts. The ratio is designated by  $k$ \replaced{(eq. \ref{eq:k} - see also \citealt{magalhaes1984}). }{ and is expressed as \citep{magalhaes1984}:} 

\added{
\begin{equation}
\label{eq:k}
k = \frac{\sum_i{C_o(W_i)}}{\sum_i{C_e(W_i)}}, 
\end{equation}}

\added{\noindent where $C_o(W_i)$ is the value of the ordinary counts in an image obtained with the retarder waveplate in the position $W_i$ and $C_e(W_i)$ is the equivalent for the extraordinary counts. A polarization measurement corresponds to 8 images, therefore $i$ runs from 1 to 8. The difference between the angular waveplate positions in two subsequent images is 22.5$\degr$. }

\added{The $k$ value can be understood as a correction for possible different efficiencies in the detection of the ordinary and extraordinary fluxes.} A value of $k$ different from the expected theoretical value can occur, for instance, if the detector sensitivity depends on the polarization of the incident beam, since the polarization of the ordinary beam is perpendicular to the polarization of the extraordinary beam. \deleted{Hence, the $k$ value can be understood as a correction for possible different efficiencies in the detection of the ordinary and extraordinary fluxes.} \deleted{It is expressed as: }

\explain{The equation that was here was moved.}

\deleted{\noindent where $C_o(W_i)$ is the value of the ordinary counts in an image obtained with the retarder waveplate in the position $W_i$ and $C_e(W_i)$ is the equivalent for the extraordinary counts. A polarization measurement corresponds to 8 images, therefore $i$ runs from 1 to 8. The difference between the angular waveplate positions in two subsequent images is 22.5$\degr$. }

In the linear polarization mode of IAGPOL, the theoretical expected value of $k$ is 1, due to the use of a half-wave retarder plate \citep[see][]{magalhaes1984}. Hence, in that case, the correction is implemented by simply multiplying the measured extraordinary counts by the $k$ value obtained from the observations. On the other hand, the quarter-wave plate is used to  measure the circular and linear polarizations. For that case, the expressions for the counts in the ordinary and extraordinary beams  given by \cite{rodrigues1998} can be used to demonstrate that the theoretical value of $k$ is a function of the Stokes parameter Q (eq. \ref{eq_k_l4}).

\begin{equation}
\label{eq_k_l4}
k = \frac{1 + 0.5 Q}{1 - 0.5 Q}.
\end{equation}

Therefore, we initially calculated the Stokes parameters considering $k$ = 1. Then we recalculated the polarization taking into account the estimated Stokes parameter Q and the resulting $k$ value was used as a multiplicative factor to the extraordinary counts. This iteration was repeated until a negligible difference in subsequent values of all Stokes parameters was achieved. An example of the application of this procedure is an absolute reduction of 0.07\% in the mean polarization error of the SW~Sex object itself. 

The above expression is valid for IAGPOL. The expression for $k$ for other polarimeters depends on the optical configuration of the  retarders and analysers of the instrument.


\bibliography{biblio}{}
\bibliographystyle{aasjournal}



\end{document}